\documentclass[hyperref={colorlinks = true,linkcolor = blue},a4paper,fleqn,anonymous]{cas-dc}

\usepackage{array}
\usepackage{makecell}
\usepackage{hyperref}
\usepackage{cleveref}
\usepackage{threeparttable}
\usepackage[authoryear]{natbib}
\usepackage{xcolor}
\def\tsc#1{\csdef{#1}{\textsc{\lowercase{#1}}\xspace}}
\tsc{WGM}
\tsc{QE}
\tsc{EP}
\tsc{PMS}
\tsc{BEC}
\tsc{DE}
\begin{document}
\let\WriteBookmarks\relax
\def\floatpagepagefraction{1}
\def\textpagefraction{.001}

%update the color into {255,0,0} to view the verion with revision highlighted in red
\definecolor{mycolor}{RGB}{0,0,0}
\definecolor{mycolor2}{RGB}{0,0,0}

% \definecolor{mycolor}{RGB}{0,0,0}
\shorttitle{\textcolor{mycolor}{Collaborative vulnerability remediation process among security technicians, users, and LLMs}}

\title[mode = title]{\textcolor{mycolor}{Practically implementing an LLM-supported collaborative vulnerability remediation process: a team-based approach}}

%  Corresponding author at: School of Information, Renmin University of China, Beijing, China.
%  E-mail addresses: wangxiaoq@ruc.edu.cn (X. Wang), tianyj@im.ac.cn (Y. Tian), keman@mit.edu (K. Huang), liangb@ruc.edu.cn (B. Liang).
%  https://doi.org/10.1016/j.cose.2024.104113
%  Received 23 February 2024; Received in revised form 31 July 2024; Accepted 8 September 2024
%  Available online 14 September 2024 
% 0167-4048/© 2024 Elsevier Ltd. All rights are reserved, including those for text and data mining, AI training, and similar technologies. 

%% Authors
\author[university]{Xiaoqing Wang}
\ead{wangxiaoq@ruc.edu.cn}
\author[university]{Yuanjing Tian}
\ead{tianyj@im.ac.cn}

\author[university,institute]{Keman Huang}
\ead{keman@mit.edu}
\corref{corresponding}
\cortext[corresponding]{Corresponding author.}

\author[university]{Bin Liang}
\ead{liangb@ruc.edu.cn}

\affiliation[university]{
organization={School of Information, Renmin University of China},
city={Beijing},
country={China}
}
\affiliation[institute]{
organization={Sloan School of Management, MIT},
addressline={Cambridge MA 02143},
country={USA}
}

\begin{abstract}
Incorporating LLM into cybersecurity operations, a typical real-world high-stakes task, is critical but non-trivial in practice.
Using cybersecurity as the study context, we conduct a three-step mix-method study to incorporate LLM into the vulnerability remediation process effectively. Specifically, we deconstruct the deficiencies in user satisfaction within the existing process (Study 1). This inspires us to design, implement, and empirically validate an LLM-supported collaborative vulnerability remediation process through a field study (Study 2). Given LLM's diverse contributions, we further investigate LLM's \textcolor{mycolor}{double-edge} roles through the analysis of remediation reports and follow-up interviews (Study 3). In essence, our contribution lies in promoting an efficient LLM-supported collaborative vulnerability remediation process.
These first-hand, real-world pieces of evidence suggest that when incorporating LLMs into practical processes, facilitating the collaborations among all associated stakeholders, reshaping LLMs' roles according to task complexity, as well as \textcolor{mycolor}{approaching the short-term side effects of improved user engagement facilitated by LLMs with a rational mindset.} 
\end{abstract}

\begin{keywords}
LLMs
\sep 
Field Study
\sep 
Mix-method
\sep 
Collaborative Vulnerability Remediation Process
\sep  
Real-world High-stakes Tasks
\sep 
Human-AI collaboration
\sep 
User Engagement

\end{keywords}

\maketitle
\section{Introduction}

The Large Language Models (LLMs), exemplified by ChatGPT \citep{ChatGPT2022} and GPT-4 \citep{openai2023gpt}, etc., have drawn intense discussions and exhibit exceptional capabilities for innovative applications ranging from creative activities \citep{petridis2023anglekindling,mirowski2023co,jones2023embodying,ashby2023personalized,wang2023popblends,hamalainen2023evaluating} to code generators \citep{jiang2021genline,jiang2022discovering,liu2023wants,mcnutt2023design,kazemitabaar2023studying}. However, integrating LLM into cybersecurity operations, a high-stakes, real-world task effectively is non-trivial \citep{wu2022ai}. \textcolor{mycolor}{First,} the potential misuse of persuasive misinformation generated by LLM \citep{zhou2023synthetic,kreps2022all,hamalainen2023evaluating}, may result into unintended consequences like catastrophic security breaches. \textcolor{mycolor}{Second, cybersecurity operations often involve complex social embedding with multi-stakeholders, including security technicians and users, making it challenging to adopt LLM into the organizational workflow} \citep{gu2021lessons,lurie2020crowdworkers,cabitza2017unintended,cai2019human,khairat2018reasons,kizilcec2016much,de2020case,van2019hiring,yang2019unremarkable,kawakami2022improving}. 

While existing efforts in the cybersecurity domain mainly focus on improving the AI tools' performance \citep{kokulu2019matched,wiczorek2019effects,ferguson2021examining,alahmadi202299}, including few studies \textcolor{mycolor}{adopting LLMs for vulnerability detection \citep{yang2024large,pearce2023examining}, automated program repair \citep{xia2023automated,huang2023empirical}, fuzzy testing \citep{deng2024large,xia2024fuzz4all,deng2023large,meng2024large}, android bug replay \citep{feng2024prompting}, } users are struggling to incorporate these advanced AIs into their daily operations \citep{iannone2022secret}. \textcolor{mycolor}{In other words, how to incorporate LLMs into the cybersecurity daily operational process and, more specifically, what LLMs can and cannot do in practice remains unclear.}

Hence, \textcolor{mycolor}{to answer this question to facilitate the effective adoption of LLMs in cybersecurity operations}, we focus on incorporating LLM into a critical cybersecurity operation task, vulnerability remediation. It refers to the process of security technicians using various tools to identify, monitor, and patch vulnerabilities in users' systems before they are exploited. It is considered high-stakes due to the substantial risks associated with potential vulnerability explorations and insufficient patching operations. \textcolor{mycolor}{More importantly, it involves both security technicians and users so that implementing LLMs into the process becomes a team-oriented collaborative task.} In particular, we start with assessing the existing vulnerability remediation process to set up the baseline, focusing on:

\begin{itemize}
    \item RQ1: What are the limitations of the existing vulnerability remediation process in practice?
\end{itemize}

\textcolor{mycolor}{Assessing the existing process to highlight its insufficient collaboration between security technicians and users can guide us in designing collaborative technology supported by LLMs to improve its performance. Inspired by the user involvement theory \citep{allen1993user,kujala2003user,bach2010involving,tarafdar2010impact,mckeen1994relationship,whyte1997understanding,haimson2023transgender}, we design the \textit{User Engagement Enhancement} and \textit{LLM-supported Technician Enhancement}, one for users and the other for security technicians, to optimize the process. The \textit{User Engagement Enhancement} component is designed to share alerts and remediation solutions with users, aiming to involve them in the remediation process. The \textit{LLM-supported Technician Enhancement} provides LLM tools for security technicians, aiming to facilitate them in remediation solution development. Hence, we conduct the second study to validate the effectiveness of this designed LLM-supported collaborative process as well as the role of each designed component.}

\begin{itemize}
    \item RQ2: Can the designed LLM-supported collaborative process effectively improve the vulnerability remediation performance? 
        \item[--]  \textcolor{mycolor}{RQ2.1: Can encouraging users to participate in the vulnerability remediation process improve the existing workflow? }
        \item[--] \textcolor{mycolor}{RQ2.2: Can additional LLM support further improve the workflow?}
\end{itemize}

Finally, \textcolor{mycolor}{to unfold LLM's contributions within this collaborative process in practice, especially the reason before the observations}, we ask the following question in Study 3:

\begin{itemize}
    \item \textcolor{mycolor}{RQ3: What role do LLMs play in the optimized collaborative process? In particular, what can and cannot LLMs do in practice, and why?}
\end{itemize}

To answer these questions, we conducted a three-step mixed-method study in a large biotechnology research institution between July 10 and August 10, 2023\footnote{Our study was exempted by our Institutional Review Board (IRB). Despite this, we still followed the same ethics and privacy requirements that an IRB would normally enforce.}. First, in Study 1, building on the DeLone and McLean Information System Success Model (D\&M model)\footnote{There also exists a rich stream of relevant studies using technology acceptance theory such as Technology Acceptance Model (TAM), Unified Theory of Acceptance and Use of Technology (UTAUT), and User Experience Theory (UX) etc. to investigate information system usage. As the D\&M model emphasizes the multidimensional aspects of the information system success and user satisfaction, which fit our goal perfectly, we adopt the D\&M model in our study. For detailed introduction of the D\&M model, UTAUT and UX theory, please refer to \citep{delone1992information,delone2003delone,jeyaraj2020delone,zheng2022ux,hornbaek2017technology,wixom2005theoretical}.}  \citep{delone1992information,delone2003delone}, which have been widely adopted in assessing user satisfaction, we survey the \textit{whole} institution and identify the insufficient user satisfaction regarding \textit{information, service, and collaboration quality} within the existing vulnerability remediation process.
 
Furthermore, in Study 2, we design the LLM-supported collaborative vulnerability process and then conduct a field study by practically implementing three distinct vulnerability remediation processes to collect empirical observations. Our data confirms that our designed process can effectively improve the process performance, including significantly promoting users' engagement, significantly reducing remediation duration, and improving user satisfaction. More specifically, \textit{User Engagement Enhancement} demonstrates the most significant contributions, including significantly increased collaboration satisfaction, shorter remediation times, and enhanced user participation. \textcolor{mycolor}{Surprisingly, the additional LLM support plays a double-edged role. On the bright side, LLMs significantly improved user participation, particularly for simpler vulnerabilities, and slightly reduced remediation times, especially for complex vulnerabilities. On the dark side, LLMs slightly increase the remediation duration for simple vulnerabilities and show limited or even negative effects on users’ satisfaction improvements.}

Taking a step further, in Study 3, by investigating the remediation reports and conducting follow-up interviews, \textcolor{mycolor}{our study reveals the dual impact of LLMs on the optimized process. Positively, it boosts user engagement by providing user-friendly guidance, code snippets, and solutions. It can also reduce remediation time for complex issues by generating quality preliminary fixes and identifying overlooked problems.
Negatively, for simple vulnerabilities, LLMs can prolong remediation due to a lack of personalized compatibility and sometimes provide impractical or inexecutable solutions, necessitating extra effort from technicians. LLMs' involvement in remediation can initially lower satisfaction as some consider vulnerability remediation out of their responsibility. However, it is worth noting that many are open to learning and following the remediation plan, with support from technicians if necessary. Hence, although engaging users may increase the remediation duration as users are not professionals like security technicians, it can ease the technicians' workload and foster a better collaborative process and cybersecurity culture. In other words, the longer remediation duration for simple vulnerabilities and a slight dip in user satisfaction can be merely a short-term hiccup and can eventually lead to long-term benefits.}

\textcolor{mycolor}{In summary, our study contributes a team-based approach to practically implement LLMs into the vulnerability remediation process, achieving a higher user engagement and lower remediation duration compared with the existing process. It also provides us with empirical evidence of the bright and dark sides of LLMs, offering valuable lessons for implementing LLMs-supported solutions in practices}. Central to these lessons is the recognition of collaboration among all stakeholders, including LLMs, to form a collaborative team as a top priority for improving high-stakes tasks. Moreover, we underscore the significance of adaptability in defining LLMs' roles, especially concerning task complexity. Last but not least, \textcolor{mycolor}{organization managers should approach the short-term side effects from improved user engagement facilitated by LLMs with a rational mindset.}

This paper's remainder is structured as follows. Section 2 reviews efforts to utilize AIs for cybersecurity operations and incorporate LLM in real-world high-stakes tasks. Section 3 examines the limitations of the existing vulnerability remediation process (Study 1). Section 4 presents the field study validating the designed LLM-supported collaborative process (Study 2). \textcolor{mycolor}{Section 5 analyzes the double-edged role of LLMs in our collaborative process (Study 3).} Finally, we conclude this study by consolidating our findings from these three studies, summarizing lessons learned from implementation in practice, and finally, pointing out limitations and future directions.

\section{Literature review}
\subsection{AI for supporting cybersecurity operations}

Cybersecurity operation, like vulnerability remediation, is a typical real-world high-stakes task that requires support from advanced AI techniques involving deep learning \citep{akgun2022new,sewak2023deep}, natural language processing  \citep{mendsaikhan2019identification,aghaei2022securebert}, knowledge representation and reasoning \citep{jia2018practical,yin2022knowledge}, as well as rule-based expert systems modeling \citep{mahdavifar2020dennes}. 

While existing efforts mainly focus on improving the AI tools' performance, the real challenge lies in effectively embedding these tools into cybersecurity operations. Some research has been conducted to incorporate AI-support tools in the cybersecurity domain \citep{jacobs2022ai,zolanvari2021trust}. For example,  \cite{jacobs2022ai} focused on synthesizing high-fidelity, low-complexity decision trees to assist network operators in detecting potential issues in network traffic with AI supports.  \cite{zolanvari2021trust} propose a universal XAI model and demonstrate its effectiveness using three cybersecurity datasets. However, trusting AI to execute cybersecurity tasks presents a dual-edged scenario \citep{taddeo2019trusting} as it potentially enables novel attacks on AI applications, posing substantial security risks. This concern can become more severe when incorporating LLMs because they are fundamentally generative models and can not distinguish security-related misconceptions effectively \citep{chen2023can}. More importantly, cybersecurity operation is a typical polyadic scenario \citep{riebe2021impact,pearlson2022design} involving collaborations among multi-stakeholders such as security technicians and users, driven by organizational workflows to avoid serious risks. This makes the integration of AIs into the operations much more challenging.

In summary, to enhance cybersecurity operations in practice, fostering polyadic collaboration among stakeholders, including security technicians, users, and AIs, is critical but challenging. This study intends to leverage LLM to effectively improve vulnerability remediation, a typical daily task in cybersecurity operations, and open gateways for further exploration in this direction.

\subsection{Large Language Models (LLMs) for real-world high-stakes tasks}

Large Language Models (LLMs), characterized by their capabilities to generate human-like text without model retraining \citep{varghese2023chatgpt}, have emerged as a transformative advancement. Their adaptability to a wide array of tasks through fine-tuning using specific datasets \citep{sejnowski2023large}, provides new ways to incorporate advanced AI into daily tasks in diverse domains \citep{sejnowski2023large,jiang2022discovering,bommasani2021opportunities}, ranging from creative activities \citep{petridis2023anglekindling,mirowski2023co,jones2023embodying,ashby2023personalized,wang2023popblends,hamalainen2023evaluating} to code generators \citep{jiang2021genline,jiang2022discovering,liu2023wants,mcnutt2023design,kazemitabaar2023studying}. 

While LLMs have proven useful, especially in low-stakes or synthesized tasks, our understanding of how to incorporate them into real-world, high-stakes tasks remains limited \citep{wu2022ai}. As LLMs are generative models developed for creating human-like texts, they can be used to generate persuasive misinformation that may not be distinguished from those human-generated ones \citep{zhou2023synthetic,kreps2022all,hamalainen2023evaluating}.
Such misinformation might just lead to minor inconveniences for those low-stakes or synthesized tasks. However, for real-world high-stakes tasks like medical decisions or cybersecurity operations, the serious consequences can manifest as incorrect diagnoses affecting downstream clinician treatment selections \citep{sivaraman2023ignore,verma2023rethinking,yang2023harnessing,tsai2021exploring,panigutti2022understanding,jacobs2021machine}, or inadequate cyber operations resulting into substantial cyber risks \citep{kokulu2019matched,wiczorek2019effects,ferguson2021examining,alahmadi202299}. 

Furthermore, many real-world tasks can be pretty complex and require multi-step interactions rather than one single run, making current LLM insufficient as an "Out-of-the-Box" solution \citep{wu2022ai,tan2021progressive,wei2022chain}.
More importantly, in practice, these tasks are typically driven by underlying processes involving various stakeholders with diverse expertise that LLMs can have unequal impacts \citep{brynjolfsson2023generative}. Hence, supporting various stakeholders as a collaborative group \citep{chiang2023two,stempfle2002thinking,huang2022being,bansal2019beyond,liang2019implicit} and delegating tasks \citep{lai2022human} among humans with varying expertise and LLMs, would be more realistic but also more challenging when adopting LLMs in real-world practices.  

In summary, while rapid developments of LLMs provide new ways to incorporate AIs for daily tasks, how to integrate them into routine cybersecurity practices is critical but remains unclear. Hence, one goal of this study is to explore how to incorporate LLM into a real-world high-stakes task, vulnerability remediation, to improve its overall performance and provide lessons that can guide collaborations with LLMs in practice.

\section{Study1: limitation of the existing vulnerability remediation process}

To achieve our goal, we conduct the first study to examine the effectiveness of the existing vulnerability remediation process, focusing on its user satisfaction in terms of the perceived information, service, and collaboration quality.

\subsection{Research design} 
\subsubsection{Existing vulnerability remediation process} 

\begin{figure*}[ht]
	\centering
		\includegraphics[scale=.45]{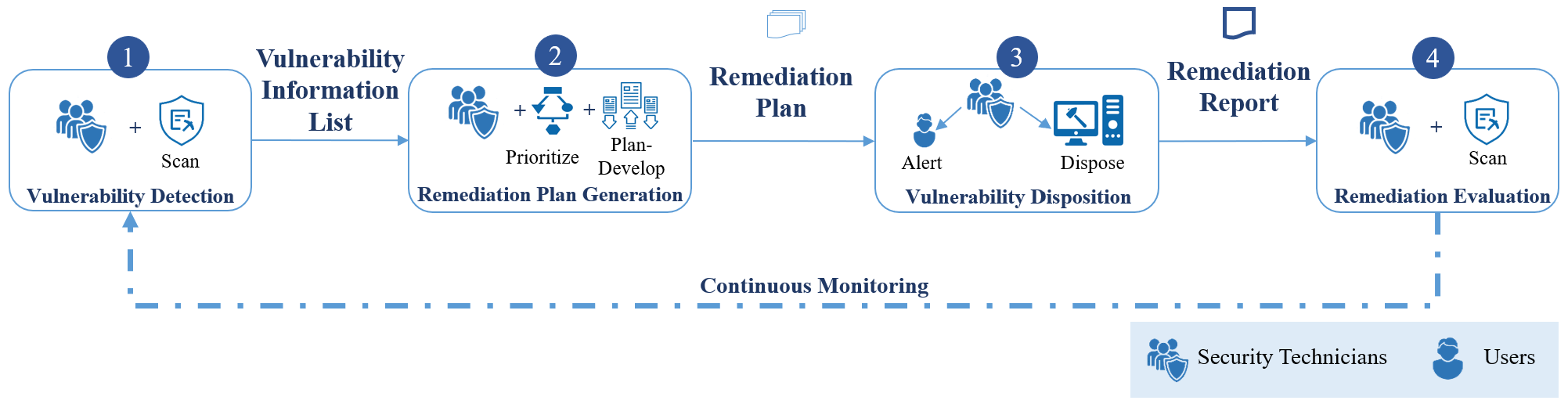}
	\caption{Existing vulnerability remediation process with four stages, including detecting the vulnerabilities, generating the remediation plan, disposing of the vulnerabilities, and reporting the remediation results.}
	\label{FIG:6}
\end{figure*}

Vulnerability remediation refers to the process in which security technicians employ various tools to identify, monitor, and patch the vulnerabilities within users' systems before they are exploited. As shown in Figure \ref{FIG:6}, during the initial stage of vulnerability detection, scanning tools automatically identify potential vulnerabilities within users' systems and gather relevant information about the vulnerable system. Subsequently, security technicians evaluate their risk levels, prioritize those higher-risk ones, and develop remediation plans. Following this, in the vulnerability disposition stage, technicians send alert information to users to schedule remediation times and execute the remediation plan to fix the vulnerabilities. Finally, in the remediation evaluation phase, after the vulnerabilities have been addressed, security technicians document the remediation results and continuously monitor the system's security status.

Notably, the vulnerability remediation tasks are considered high-stakes, due to the substantial risks associated with potential vulnerability explorations and insufficient patching operations. Furthermore, it requires the collaboration of security technicians and users in nature because users own the vulnerable devices, and the security technicians are tasked to remediate the identified vulnerabilities. This makes vulnerability remediation naturally a collaborative and high-stakes task, aligning with the context of our study. 

\subsubsection{Theoretical background: D\&M model}
To examine the effectiveness of the existing process, we adopt the DeLone and McLean (D\&M) information system success model, which is one dominant framework designed to assess the performance of information systems. Originally proposed in 1992 \citep{delone1992information} and later updated in 2003 \citep{delone2003delone}, the model posits a strong relationship between information systems success and user satisfaction. More specifically, user satisfaction is a key indicator of an information system's success \citep{jeyaraj2020delone} and is directly influenced by perceived \textit{System Quality}, \textit{Information Quality}, and \textit{Service Quality}. As collaboration and knowledge sharing among multi-stakeholders become increasingly critical for advanced information systems ( i.e. enterprise portals, social media, and e-learning etc.), subsequent studies \citep{urbach2010improving,cidral2018learning,victor2013process,white2018help,ziegele2018dynamics,marek2015evaluating} further extend the D\&M model by incorporating the \textit{Collaboration Quality} as an essential dimension for user satisfaction, which evaluates the effectiveness of the integration and exchange of information, decision-making, and resource sharing among different stakeholders \citep{johnson2002business}.

Overall, the rationale of the D\&M model is that a high-quality system that provides accurate and timely information, supported by excellent service with smooth collaboration among multi-stakeholders, will likely lead to greater satisfaction. This satisfaction, in turn, contributes to the achievement of the system's net benefits, which is the overall value of the system to its users and underlying organizations. Thus, in the D\&M model, user satisfaction is both an outcome measure of success and a mediator that influences the overall effectiveness and utility of an information system.

\subsubsection{Measurement of user satisfaction}

As suggested by \cite{jeyaraj2020delone}, user satisfaction is more frequently chosen as the critical dimension for information system success. Hence, we focus on user satisfaction to proxy the effectiveness of the existing vulnerability remediation process and develop our measures based on the D\&M model. In particular, as vulnerability remediation is a collaborative process rather than a system that users and security technicians can use, we exclude the \textit{System Quality} dimension. As reported in Table \ref{tab:construct}, we adopt the measurements from previous studies. Then, two researchers revised them to match our research context independently, followed by a group discussion involving the entire research team, including the Chief Security Officer in the institution, to avoid confusion from sentence expressions. Additionally, we use a 5-point Likert scale design for each item, starting from a scale of strongly disagree (1) to strongly agree (5).

\begin{itemize}
    \item Information Quality. Following previous studies \citep{vitturi2013industrial,delone2003delone,bailey1983development,hartono2005analisis,iivari2005empirical}, we consider users' perceived timeliness, accuracy, and consistency of the provided \textcolor{mycolor2}{alerting} information to \textcolor{mycolor2}{assess} the information quality.
    \item Service Quality. We measure users' perceived responsiveness and reliability of the remediation supports provided by the security technicians \citep{victor2013process,delone2003delone,marek2015evaluating,mohammed2022evaluating}.
    \item  Collaboration Quality. We proxy the collaboration quality using help-seeking willingness \citep{marek2015evaluating,bailey1983development,white2018help} and involvement willingness \citep{marek2015evaluating,bailey1983development,ziegele2018dynamics} which represents users' willingness to seek help from security technicians and be involved in the vulnerability remediation process.
\end{itemize}

\begin{table*}[!htbp] 
\normalsize
\caption{Construct Definitions and Items to Measure Effectiveness of Vulnerability Remediation Process.}
\label{tab:construct}
\resizebox{\linewidth}{!}{
\begin{tabular}{@{}p{1.6cm}p{3.6cm}p{5cm}p{6cm}p{5cm}@{}}
\toprule
Construct & Measure &
  Definition &
  Item &
 Source \\ \midrule
\multirow{3}{3cm}[+1ex]{Information \\ Quality} &
 Information Timeliness &
  The timeliness of alerting information in resolving cybersecurity incidents. &
  I believe that alerting information is sent very promptly and generally arrives before a system compromise. &
\cite{vitturi2013industrial,delone2003delone,bailey1983development,hartono2005analisis}
  \\ 
 &
  Information Accuracy &
  The accuracy of alerting information in resolving cybersecurity incidents. &
  I believe that alerting information is very accurate and can help me effectively resolve cybersecurity incidents. &
\cite{ghazi2020private,delone2003delone,bailey1983development,hartono2005analisis}
  \\  
 &
  Information Consistency &
  The consistency of existing alerting information delivery methods with user expectations. &
  The way to deliver cybersecurity incident alerting information is consistent with what I expect. &
  \cite{iivari2005empirical}
  \\ \midrule

\multirow{2}{3cm}[+0ex]{Service 
\\  
Quality} 
   
   &
   Assistance Responsiveness &
  Users' perception of the timeliness of assistance from security technicians. &
  After receiving cybersecurity alerting information, assistance from security technicians can usually resolve cybersecurity issues promptly. &
  \cite{victor2013process,delone2003delone,mohammed2022evaluating}
  \\
   &
  Assistance Reliability &
  Users' perception of the utility of assistance from security technicians. &
  After receiving cybersecurity alerting information, assistance from security technicians can usually resolve cybersecurity issues thoroughly. &
\cite{victor2013process,delone2003delone,mohammed2022evaluating}
\\ \midrule

\multirow{2}{3cm}[+0ex]{Collaboration
\\  
Quality}  &
  Help-seeking Willingness &
  Users' inclination to seek assistance from security technicians after receiving cybersecurity alerting information. &
  After receiving cybersecurity alerting information, I am willing to actively seek assistance from security technicians rather than handle it myself. & \cite{bailey1983development, white2018help, marek2015evaluating}
  \\  
   &
  Involvement Willingness &
  Users' willingness to collaboratively engage in the vulnerability remediation process and learn from security technicians. &
   If security technicians teach cybersecurity problem-solving methods while handling cybersecurity issues, I am willing to learn. & \cite{bailey1983development, ziegele2018dynamics, marek2015evaluating}

  \\ \bottomrule
\end{tabular}
}
\end{table*}

% Note that not all users within the institution have ever been involved in the vulnerability remediation process before we conducted this study. Hence, rather than asking users' satisfaction on the existing vulnerability remediation process directly, we design a question to separate them into two groups, one who received at least one alerting information and was involved in the vulnerability remediation process before, while the other never did so: 

% 请注意，在我们进行这项研究之前，并非机构内的所有用户都参与过漏洞修复过程。因此，我们没有直接询问用户对现有漏洞修复过程的满意度，而是设计了一个问题，将他们分为两组，一组收到至少一个警报信息并参与过漏洞修复过程，另一组从未参与过：

Note that not all users within the institution have ever been involved in the vulnerability remediation process before we conducted this study.
\textcolor{mycolor2}{If we only include users with experience in vulnerability remediation, we can obtain their experience regarding information quality, service quality, and collaboration quality of the existing process. However, it is impractical to assess whether experiencing the existing vulnerability remediation process increases or decreases users’ satisfaction. Therefore, we design a question to separate users into two groups, one who received at least one alerting information and was involved in the vulnerability remediation process before, while the other never did so: 
}
% 如果只包含有漏洞修复经验的用户，只能获得现有漏洞修复流程information quality ; service quality ; collaboration quality相关指标的绝对值，但难以对这些绝对值的优劣做出客观评价。

\begin{itemize}
    \item \textbf{Group without previous vulnerability remediation experience ($G_{w/o}$).} This group has not been involved in the vulnerability remediation process before. 
    \item \textbf{Group with previous vulnerability remediation experience ($G_w$).} This group received at least one alerting information and was involved in the vulnerability remediation process before. 
\end{itemize}

\textcolor{mycolor2}{Hence, the data from Group $G_{w/o}$ reveals users' \textit{expectation} on the existing process, while the data from Group $G_w$ represents users' satisfaction from their \textit{actual} experience. The difference between these two groups will reveal whether the previous experience is positive or negative. If the experience with the existing process is positive, we would expect a higher user satisfaction for $G_w$ compared with $G_{w/o}$. Otherwise, the negative experience will make the measure in $G_w$ lower.}
In other words, our goal is not to investigate the relations among these constructs discussed above to validate the D\&M model. Instead, we aim to employ the D\&M model to uncover the limitations within the existing vulnerability remediation process.

% Hence, the data from Group $G_{w/o}$ reveals users' \textit{expectation} on the existing process, while the data from Group $G_w$ represents users' satisfaction from their \textit{actual} experience. Therefore, the difference between these two groups will reveal whether the previous collaboration are positive or negative. If the experience with the existing process is positive, we would expect a higher user satisfaction for $G_w$ compared with $G_{w/o}$. Otherwise, the negative collaboration will make the measure in $G_w$ lower. In other words, our goal is not to investigate the relations among these constructs discussed above to validate the D\&M model. Instead, we aim to employ the D\&M model to uncover the limitations within the existing vulnerability remediation process.

% 如果只包含有漏洞修复经验的用户，只能获得这部分用户对现有漏洞修复流程information quality ; service quality ; collaboration quality的实际体验数据。然而由于缺少对这三种quality的客观评估指标（评级体系），因此仅通过实际体验数据评估现有漏洞修复流程是不现实的。

% 因此，将没有漏洞修复经验的用户纳入研究，他们对现有漏洞修复流程information quality ; service quality ; collaboration quality的评价代表了用户对漏洞修复流程的期待。

% 总的来说，来自组$G_{w/o}$的数据揭示了用户对现有流程的\textit{期望}，而来自组$G.w$的数据代表了用户对\textit{actual}体验的满意度。因此，这两组之间的差异将揭示之前的体验是积极的还是消极的。
% 研究结果发现，来自组$G_{w/o}$的数据 小于 来自组$G_{w/o}$的数据，说明用户对现有漏洞修复流程的体验是消极的，是不够满意的。通过对比这种方式实现了对现有流程的评估。

\subsection{Participant demographics}

We distributed our questionnaires to \textit{all} the employees and students, around 600 in total, in the institution between July 10 and July 16, 2023. The questionnaires are designed to be anonymous without collecting personal identification. We inform participants of the goal of this survey, and they can fill out the questionnaires voluntarily. We also included an attention-checking question (i.e., questions in which participants were asked to answer the official name of the institution) so that we could filter out those inattentive participants. 266 questionnaires were completed, with a 44.33\% response rate. We exclude those who did not pass our attention checks and those who chose the first option for all scale items. As we focus on user satisfaction, we also exclude responses from security technicians, leaving us with 246 valid responses for further analysis.
This satisfies the sample size requirement estimated using the Raosoft Sample Size Calculator\footnote{http://www.raosoft.com/samplesize.html}: with an error rate of 5\%, confidence of 95\%, a response distribution of 50\% among a population of 600 people, the preferred sample size of the survey should be more than 235.

\begin{table*}[hbtp]
\caption{Participant Demographics in Study1.}
\label{tab:participant}
\begin{tabular}{llll}
\toprule
\textbf{Item}          &                    & \textbf{Frequency (n) }& \textbf{Percentage (\%)} \\ 
\midrule
Gender    & Female    & 150          & 60.98    \\
          & Male      & 96          & 39.02    \\
Age     & 26-35         & 92            & 37.40            \\
        & 36-45         & 102           & 41.46            \\
        & 45-55         & 40           & 16.26           \\
        & $\geq$ 55      & 12          & 4.88            \\
Job Category  & Employees    & 216          & 87.80    \\
          & Students     & 30          & 12.20    \\
Job Responsibility     &Research    & 159     &64.63    \\
        &Technical    & 56           & 22.76           \\
        &Management  & 31           & 12.60           \\
Job Role  &Regular User      & 156          & 63.41     \\
            &Departmental Security Officer  & 63           & 25.61           \\
        &Departmental Security Manager    & 27           & 10.98    \\

              \bottomrule
\end{tabular}
\end{table*}

As reported in Table \ref{tab:participant}, the participant profile shows a diverse group with a majority (60.98\%) of female respondents. In terms of age distribution, participants primarily fall within the age brackets of 26-45 years, with 37.40\% in the 26-35 age range and 41.46\% in the 36-45 age range. There is also representation from older age groups, with 16.26\% aged between 46-55 years and 4.88\% aged over 55 years. For job categories, the majority (87.80\%) of participants are employees, while students make up the remaining 12.20\%. For job responsibility, the data highlights a significant presence in research positions, constituting 64.63\% of the participants, followed by technical support positions at 22.76\%, and a smaller representation in management positions at 12.60\%. Additionally, the majority of participants identify as regular users (63.41\%), followed by departmental security officers (25.61\%), and a smaller portion as departmental security managers (10.98\%).\footnote{Note that departmental security officers and managers are considered users rather than security technicians in our context. This is because they mainly serve in part-time roles within each department for liaising with security matters and collaborating with security technicians to patch vulnerabilities.} Finally, among these 246 participants, 186 report previous experiences belonging to group $G_w$. The other 60 are within group $G_{w/o}$ without being involved in vulnerability remediation before. 

\subsection{Results: insufficient user satisfaction within existing vulnerability remediation process}

In this section, we report our observation focusing on \textit{whether experiencing the existing vulnerability remediation process will result in lower user satisfaction}. More specifically, we use a one-tailed student t-test or Mann-Whiteney U-test depending on whether the data follows the normal distribution. Once the result is significant, we will report the p-value ($p$) and its effect size ($d$) \footnote{We will follow the same protocol to analyze the data and report results in Study 2.}.

\begin{figure*}[ht]
	\centering
		\includegraphics[scale=.36]{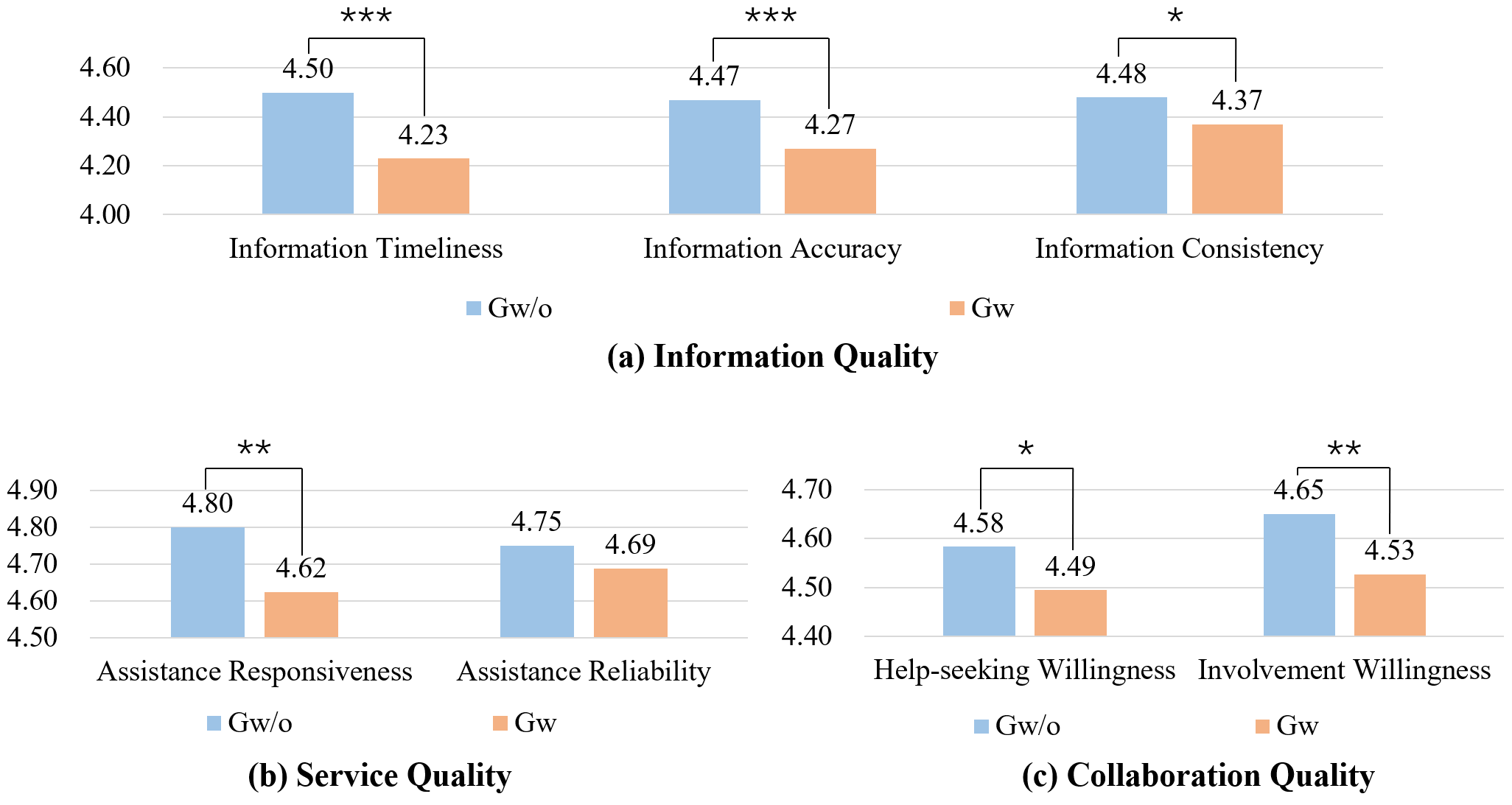}
	\caption{Users' satisfaction with information quality, service quality, and collaboration quality in Study1. \textit{*p<0.1,**p<0.05,***p<0.01.}}
	\label{FIG:study1result}
\end{figure*}

\subsubsection{Information quality}
As summarized in Figure \ref{FIG:study1result}(a), users from $G_{w}$ group who have experienced vulnerability remediation before report a significantly lower value regarding information' timeliness (\textit{p = 0.006, d = 0.330}), accuracy (\textit{p = 0.009, d = 0.249}) and consistency (\textit{p = 0.050, d = 0.141}). 
This suggests that users who have received alerting information regarding the related vulnerability are not satisfied with their quality regarding timeliness, accuracy, and consistency. Hence, improving the information quality is one direction for optimizing the existing process.

\subsubsection{Service quality}
As reported in Figure \ref{FIG:study1result}(b),  we observe a consistently lower value for group $G_{w}$, confirming a negative service experience within the existing process. In particular, users who have experienced the existing process consider the support from security technicians not responsive enough ($p=0.016, d=0.276$). This suggests improving the security technician's capability to respond to users' requests more timely.

\subsubsection{Collaboration quality}
As illustrated in Figure \ref{FIG:study1result} (c), users who have experienced the existing process report a significantly lower willingness to seek assistance from security technicians ($p=0.051, d=0.117$) and engage with security technicians ($p=0.047, d=0.183$).  In other words, the previous collaboration quality in vulnerability remediation demonstrates a negative impact, highlighting the necessity to make the collaboration among users and security technicians more effective.

\textbf{In summary, the user satisfaction with the existing vulnerability remediation is consistently insufficient regarding the information, service, and collaboration quality. Hence, to optimize the existing process, we should design mechanisms to improve these aspects, which is what our study aims for.}

\section{Study2: LLM-supported collaborative vulnerability remediation process}

Our Study 1 uncovered that the existing process has insufficient user satisfaction in terms of information, service, and collaboration quality. Therefore, inspired by the user involvement theory  (i.e. \citep{allen1993user,kujala2003user,bach2010involving,tarafdar2010impact,mckeen1994relationship,whyte1997understanding,haimson2023transgender}), Study 2 seeks to enhance this process by designing two components, \textcolor{mycolor}{the user engagement enhancement for users and the LLM-supported technician enhancement for security technicians},
and then conducting a field study to validate their effectiveness.

\subsection{Process design}

\textcolor{mycolor}{As suggested by the user involvement theory (i.e. \citep{allen1993user,kujala2003user,bach2010involving,tarafdar2010impact,mckeen1994relationship,whyte1997understanding,haimson2023transgender}), involving users is crucial for user satisfaction and the success of information systems. It can make the system better understood, perceived to be useful, appropriately configured, and more valued and better accepted by users. In particular, involving users can lead to higher user information satisfaction and improve their confidence in utilizing the provided information \citep{tarafdar2010impact}. It also leads to greater communication and collaboration between users and information systems professionals (security technicians in our study context)\citep{tarafdar2010impact,mckeen1994relationship}, as well as enables users to present and clarify their needs \citep{tarafdar2010impact,whyte1997understanding}, resulting in higher perceived collaboration and service quality. Recent studies on polyadic human-AI interactions also highlight enhancing communication and engagement within multi-user collaborative teams to improve their overall team performance \citep{zheng2022ux}. However, as elaborated in Figure \ref{FIG:6}, when taking a closer look at the existing vulnerability remediation process, the collaboration between technicians and users is limited: users are rarely involved in the vulnerability remediation task, and security technicians take almost all responsibilities. This can be the reason for the insufficient user satisfaction revealed in Study 1.} Hence, this motivates us to design the first component \textit{User Engagement Enhancement}, which shares alert information and remediation solutions with users, aiming at involving them in the vulnerability remediation process:

\begin{itemize}
    \item \textbf{User Engagement Enhancement.} As illustrated by the red dashed line in Figure \ref{FIG:design}(b) and (c), this collaborative mechanism simultaneously delivers alert information to security technicians and users. More importantly, users can contact security technicians for support, and technicians can also share the remediation plan with users so that users can engage in the vulnerability remediation process more actively.
\end{itemize}

\textcolor{mycolor}{Furthermore, user involvement alone may not be sufficient to achieve significant performance improvement (\cite{he2008role,kujala2003user}). The effects of user involvement on user satisfaction and information system success can be constrained by various situational factors such as task complexity, user knowledge, and type of participants (\cite{edstrom1977user}). In particular, Ives and Olson's cognitive framework (\cite{ives1984user}), along with subsequent studies (i.e.\cite{choe1996relationships,curtis1988field,he2008role}) highlights the importance of reducing users' cognitive burden to achieve positive user involvement outcomes. Therefore,} given the advantage of LLM models, on top of the \textit{User Engagement Enhancement}, we further design the \textit{LLM-supported Technician Enhancement}. This component provides LLM \footnote{Before implementing our process, we explored several available cutting-edge Large Language Model (LLM) tools, including ChatGPT (GPT-3.5), Chat8 (GPT-3.5), and Monica (GPT-4). Among these tools, Monica, powered by GPT-4 \citep{openai2023gpt}, demonstrated sufficient capability for vulnerability remediation plan generation. Hence, we utilized Monica in our study. However, our preference for Monica primarily stems from the advanced capabilities provided by GPT-4. To avoid confusion, in our manuscript, we consistently refer to the adopted tool as \textit{"LLM"}, rather than "Monica". } for security technicians, \textcolor{mycolor}{aiming at facilitating them} to develop a more comprehensive and user-friendly remediation plan.

\begin{itemize}
    \item \textbf{LLM-supported Technician Enhancement.}
    As illustrated by the black dotted line in Figure \ref{FIG:design}(c), this mechanism further empowers security technicians with LLM when developing and refining the vulnerability remediation plan to engage users in vulnerability remediation.
\end{itemize}

It is worth noting that in this optimized process, LLM is only provided to security technicians, not users. This is because fixing vulnerabilities requires specialized expertise. However, users in our study did not have such vulnerability remediation knowledge and had not used LLM for such high-stakes tasks before. Hence, to avoid bringing too much cognitive burden on users, which may harm the effectiveness of the process, we did not introduce users to LLM. Instead, they will receive the remediation solution from security technicians, which was developed with support from LLM. Once the solution is user-friendly enough, they can follow the solution without specialized security expertise requirements. However, customizing an LLM tool that is friendly enough for users can be valuable to improve the process's performance further.

\subsection{The field study to examine the effectiveness of designed process}

\subsubsection{Conditional design}

\begin{figure*}[ht]
	\centering
		\includegraphics[scale=.5]{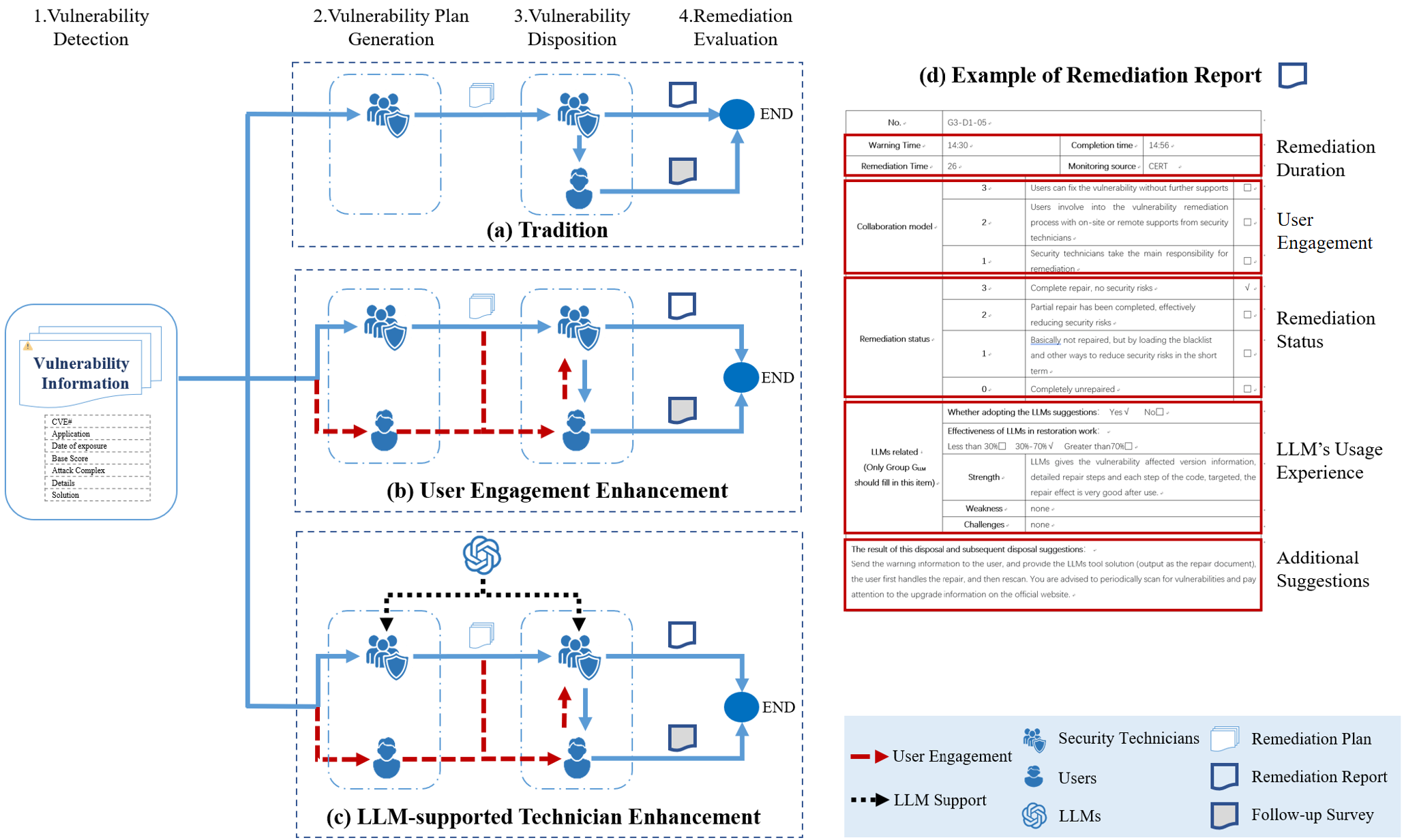}
	\caption{Experimental conditions and process designs for the three different vulnerability remediation processes. The red dashed lines and black dotted lines within \textit{User Engagement Enhancement} and \textit{LLM-supported Technicians Enhancement} conditions represent the logical workflow for the two designed components, respectively. The right part is an example of the remediation report that security technicians must submit after patching a vulnerability.}
	\label{FIG:design}
\end{figure*}

To examine the effectiveness of our LLM-supported collaborative vulnerability remediation process, we conducted a field study in the institution from July 17 to 21 and July 24 to 28, 2023. More specifically, as reported in Figure \ref{FIG:design}, we practically implement and run three different vulnerability remediation processes as follows\footnote{Due to the limited resources to manage the field study, the experiments for $G_{ta}$ and $G_{ue}$ were conducted simultaneously from July 17th to 21st, while the experiments for $G_{llm}$ were conducted from July 24th to 28th. Notably, these three experiments are operated independently. Additionally, the experiments were paused on July 22nd and 23rd, 2023 as these days are the weekend.}: 

\begin{itemize}
    \item \textbf{Tradition ($G_{ta}$)}. In this case, as shown in Figure \ref{FIG:design}(a), the security technicians and users follow the existing vulnerability remediation process described in Section 3.1.
    \item \textbf{User Engagement Enhancement ($G_{ue}$)}. In this case, as shown in Figure \ref{FIG:design}(b), unlike the \textit{Tradition} case, the alerting information with the identified vulnerability information will be shared with users simultaneously, and security technicians will provide remediation solutions to users. Therefore, based on how users participate in the vulnerability remediation process, there exist three security technicians-users collaboration models:
        \begin{itemize}
            \item \textit{\textbf{User Mainly}} where users fix the vulnerability following the provided remediation solution with minimal or no support from security technicians.
            \item \textit{\textbf{Technicians \& Users}} where users are involved in the vulnerability remediation process with on-site or remote support from security technicians and they collectively patch the vulnerabilities.
            \item \textit{\textbf{Technicians Mainly}} where security technicians take the main responsibility for remediation.
        \end{itemize}
    \item \textbf{LLM-supported Technicians Enhancement ($G_{llm}$)}.  In this case, as shown in Figure \ref{FIG:design}(c), on top of the \textit{User Engagement Enhancement}, LLM provides additional support to security technicians for vulnerability remediation solution generation and adjustment.
\end{itemize}

\textcolor{mycolor}{It is worth noting that we did not consider the condition where LLM is provided to security technicians without the user engagement enhancement component. On the one hand, inspired by the user involvement theory, as well as the group discussions involving the Chief Security Officer in the institution, the user engagement enhancement component is critical to improving the process into a collaborative process, while the LLMs were expected to further improve its effectiveness. On the other hand, like other cybersecurity operational processes, the vulnerability remediation process involves both security technicians and users. Hence, we are interested in the role of LLMs in such a polyadic human-AI collaboration scenario rather than merely using LLMs as a tool for security technicians. As we will report later in Section 4.3, while confirming the expected effectiveness of the user engagement component, we observe a complex heterogeneous effect of the additional LLM support for security technicians. This motivates us to conduct a further study, Study 3, through remediation reports analysis and follow-up interviews to investigate LLMs' double-edge effects in practice.}

\subsubsection{Vulnerability selection}

Before starting our field study, we used the scanning tool to scan the entire IT environment within the institution to identify user system vulnerabilities. We only focus on high-risk vulnerabilities presented in at least three different users' systems, resulting in 29 high-risk vulnerabilities (see Table \ref{tab:vulnerability}) as remediation tasks for our study. More specifically, based on the remediation complexity, these vulnerabilities belong to two groups:

\begin{table*}[!ht]
\caption{Selected Vulnerability Information in Study2.}
\label{tab:vulnerability}
\resizebox{\linewidth}{!}{
\begin{tabular}{@{}cccc@{}}
\toprule
Number & Vulnerability Name                                                                & Category     & \makecell{   Complexity} \\ \midrule
T1     & Apache   HTTP Server Authorization Issue Vulnerability (CVE-2022-31813)           & Mysql/Apache & Complex                  \\
T2     & Apache   HTTP Server Buffer Overflow Vulnerability (CVE-2021-44790)               & Mysql/Apache & Complex                  \\
T3     & Apache   HTTP Server Buffer Error Vulnerability (CVE-2021-39275)                  & Mysql/Apache & Complex                  \\
T4  & Apache   HTTP Server ap\_get\_basic\_auth\_pw Authentication Bypass Vulnerability   (CVE-2017-3167) & Mysql/Apache & Complex \\
T5     & Apache   HTTP Server mod\_mime Buffer Overflow Vulnerability (CVE-2017-7679)      & Mysql/Apache & Complex                  \\
T6     & Oracle   MySQL Server/MariaDB Security Vulnerability (CVE-2016-9843)              & Mysql/Apache & Complex                  \\
T7     & Apache   HTTP Server Input Validation Error Vulnerability (CVE-2022-22721)        & Mysql/Apache & Complex                  \\
T8  & Oracle   MySQL cURL Component Input Validation Error Vulnerability (CVE-2022-27778)                 & Mysql/Apache & Complex \\
T9  & Oracle   MySQL/MariaDB Packaging (OpenSSL) Full Vulnerability (CVE-2022-0778)                       & Mysql/Apache & Complex \\
T10    & Target   Host showmount -e Information Disclosure (CVE-1999-0554)                 & Scarce/Rare  & Complex                  \\
T11    & SNMP   Service  Writable Community String (CVE-1999-0516)                & Scarce/Rare  & Complex                  \\
T12    & NTP Mode 6 Detection   Vulnerability                                              & Scarce/Rare  & Complex                  \\
T13    & OpenSSL   Operating System Command Injection Vulnerability (CVE-2022-1292)        & SSL/SSH      & Simple                   \\
T14 & Oracle   MySQL OpenSSL Component Input Validation Error Vulnerability (CVE-2022-1292)               & SSL/SSH      & Simple  \\
T15    & OpenSSH   MaxAuthTries Restriction Bypass Vulnerability (CVE-2015-5600)           & SSL/SSH      & Simple                   \\
T16    & SSL   Weak Encryption Algorithms                                                  & SSL/SSH      & Simple                   \\
T17    & OpenSSH   Command Injection Vulnerability (CVE-2020-15778)                        & SSL/SSH      & Simple                   \\
T18    & OpenSSH   Security Restriction Bypass Vulnerability (CVE-2016-10012)              & SSL/SSH      & Simple                   \\
T19    & OpenSSH   Input Validation Error Vulnerability (CVE-2019-16905)                   & SSL/SSH      & Simple                   \\
T20 & OpenSSH do\_setup\_env   Function Privilege Escalation Vulnerability (CVE-2015-8325)                & SSL/SSH      & Simple  \\
T21    & SSL/TLS   Protocol Information Disclosure Vulnerability (CVE-2016-2183)           & SSL/SSH      & Simple                   \\
T22    & SSL   Medium Strength Encryption Algorithms (CVE-2016-2183)                       & SSL/SSH      & Simple                   \\
T23    & OpenSSH   auth\_password Function Denial of Service Vulnerability (CVE-2016-6515) & SSL/SSH      & Simple                   \\
T24    & OpenSSH   Remote Code Execution Vulnerability (CVE-2016-10009)                    & SSL/SSH      & Simple                   \\
T25    & OpenSSH   Security Vulnerability (CVE-2021-41617)                                 & SSL/SSH      & Simple                   \\
T26    & OpenSSH   Information Disclosure Vulnerability (CVE-2020-14145)                   & SSL/SSH      & Simple                   \\
T27    & SSL/TLS   RC4 Information Disclosure Vulnerability (CVE-2013-2566)                & SSL/SSH      & Simple                   \\
T28    & SSL/TLS Bar-Mitzvah   Attack Vulnerability (CVE-2015-2808)                        & SSL/SSH      & Simple                   \\
T29    & Server   Supports TLS Client-initiated Renegotiation Attack (CVE-2011-1473)       & SSL/SSH      & Simple                   \\ \bottomrule
\end{tabular}
}
\end{table*}

\begin{itemize}
    \item \textit{\textbf{Complex}}. This group includes vulnerabilities related to Mysql, Apache, and three less common ones. Mysql and Apache are widely used for end-user services. Patching their associated vulnerabilities involves coordinating low-traffic periods to reduce impacts, checking historical operations for compatibility, and backing up databases to avoid unintended consequences, etc. The three uncommon vulnerabilities are rarely found and lack easy, readily available remediation solutions, so developing a remediation solution from scratch could require significant effort. In other words, patching these vulnerabilities is relatively complicated.
    \item \textit{\textbf{Simple}}. The vulnerabilities within this group are all SSL/SSH-related vulnerabilities. Both OpenSSL and OpenSSH are communication layer software. Patching these vulnerabilities will not impact the application and data access deployed above them. Hence, their remediation is expected to be relatively simpler.
\end{itemize}

Hence, the difference in complexity required to patch these two groups of vulnerabilities will enable us to investigate how the task complexity can affect the process's effectiveness.

\subsubsection{Participant selection}
\textcolor{mycolor2}{For each selected vulnerability, we randomly select three users whose devices had that specific vulnerability and then randomly assign them to the three conditions.} If a user's system contains multiple selected vulnerabilities, we involve her/him only once. Hence, each condition involves 29 users, resulting in 87 participants in our field study. Additionally, following the industrial best practices, two security technicians responsible for vulnerability remediation in the institution are assigned to each condition to lead the process. They collaborate to patch 29 vulnerabilities (12 complex and 17 simple) with 29 associated users. Specifically, the two security technicians for condition $G_{llm}$ had some prior LLM usage experiences. Thus, we can mitigate the potential negative impacts due to lack of experience. Notably, as supporting vulnerability remediation is part of users' job responsibility and our study did not collect any personal information, users were deliberately kept uninformed about the purpose of this study to minimize potential bias stemming from knowledge of the research objectives\footnote{The vulnerability remediation process only collects public information on vulnerable devices like system version and configuration which is necessary for remediation solution, and we did not collect any personal identical information. It is also a routine workflow in the institution we select so that our study does not pose any additional privacy concerns. Hence, our study was exempted by our Institutional Review Board (IRB).}. However, we inform security technicians of the goal and request them not to communicate with those assigned to other conditions to keep these processes independent. Overall, our study involves 87 users and 6 security technicians in three independent processes.

\subsubsection{Experimental process}

Before starting the experiment, following the vulnerability and participant selection process described above, we identified 29 vulnerabilities and assigned the associated 29 users and two security technicians for each condition. Then, throughout our field study, on a daily basis, we randomly select the vulnerabilities to initiate the remediation processes and involve associated users and security technicians in vulnerability remediation plan generation and patching. Once the vulnerabilities are fixed, \textcolor{mycolor}{as shown in Figure \ref{FIG:design} (d),} the security technicians must submit a remediation report for each vulnerability \textcolor{mycolor}{which includes remediation duration, user engagement level, evaluation of technicians' experience with LLMs (only in $G_{llm}$). Finally, }users are invited to complete a post-survey using the same questionnaires from Study 1. \textcolor{mycolor}{In other words, we collect feedback from both security technicians and users regarding each vulnerability remediation, enabling us to investigate the effectiveness of our designed collaborative process.}

\subsubsection{Measurement}

Beyond measuring user satisfaction regarding users' perceived \textit{Information}, \textit{Service} and \textit{Collaboration Quality} as Study 1, \textcolor{mycolor}{using security technicians' feedback through remediation reports as shown in Figure 3(d)}, we develop the following measures to evaluate the \textcolor{mycolor}{user engagement level and remediation duration}, with a particular focus on the variations observed across different conditions ($G_{ta}$, $G_{ue}$ and $G_{llm}$).

\textbf{User engagement level} is an objective metric that reflects the extent to which users are actively involved in the remediation process \citep{muresan2019chats,mitchell2021automated}.
In both conditions, $G_{ue}$ and $G_{llm}$, users decide how to participate in the vulnerability remediation process. Hence, we can measure the user engagement level ($ue(v_k, G_{i}), G_{i} \in \{G_{ue}, G_{llm}\}$) based on the collaborative model \textcolor{mycolor}{(see Section 4.2.1 for more details)} of their participation, as follows:
% collaborative model is
% \begin{small} 

\begin{equation}
ue(v_k,G_{i}) =  \begin{cases}
  1 & \text{ if  } \text{\textit{Technicians Mainly} remediate} \\
  2 & \text{ if  } \text{\textit{Technicians \& Users} remediate} \\
  3  & \text{ if  } \text{\textit{Users Mainly} remediate}
\end{cases}
\label{equation-ue}
\end{equation}

% \end{small} 

where a higher $ue(v_k,G_{i})$ indicates a higher user engagement level when patching vulnerability $v_k$. Then we can measure the user engagement level change $UEC(v_k,G_{i},G_{j})$ for each vulnerability $v_k$ between two conditions as:

\begin{equation}
UEC(v_k,G_{i},G_{j}) =  \begin{cases}
  1 & \text{ if  }  ue(v_k,G_{i}) > ue(v_k,G_{j}) \\
  0 & \text{ if  }  ue(v_k,G_{i}) = ue(v_k,G_{j}) \\
  -1  & \text{ if  }  ue(v_k,G_{i}) < ue(v_k,G_{j})
\end{cases}
\label{equation-uec}
\end{equation}

Therefore, a significantly positive user engagement level change when comparing condition $G_{i}$ and condition $G_{j}$, a.k.a. $UEC(G_{i},G_{j})$ indicates that the user engagement level in $G_{i}$ is higher than $G_{j}$.

\textbf{Remediation duration}, the time spent on patching a vulnerability, is an objective measure for process efficiency \citep{kumar2014triggering}. As shown in Figure \ref{FIG:design}(d), the security technicians will report the time receiving the alerting information and the time finishing remediation. Hence, we develop the \textit{remediation duration change ($RDC$)} to investigate whether our designed process can effectively reduce the remediation duration. For each vulnerability $v_k$, we consider:

\begin{equation}
RDC(v_k,G_{i},G_{j}) =  \begin{cases}
  1 & \text{ if  }  rd(v_k,G_{i}) > rd(v_k,G_{j}) \\
  0 & \text{ if  }  rd(v_k,G_{i}) = rd(v_k,G_{j}) \\
  -1  & \text{ if  }  rd(v_k,G_{i}) < rd(v_k,G_{j})
\end{cases}
\label{equation-c}
\end{equation}

where $rd(v_k,G_{i})$ represents the remediation duration for vulnerability $v_k$ in condition $G_{i} \in \{G_{ta},G_{ue},G_{llm}\}$. For example, if a vulnerability requires a lower remediation time duration in condition $G_{ue}$ compared with $G_{ta}$, we will assign its remediation duration change $RDC(v_k, G_{ue}, G_{ta})$ as $-1$, otherwise, $1$ will be assigned. Hence, if we observe a significantly negative remediation duration change when comparing $G_{ue}$ with $G_{ta}$, $a.k.a.$ $RDC(G_{ue}, G_{ta})$, we can conclude that the remediation duration time in condition $G_{ue}$ is reduced.

\subsection{Results: the effectiveness of designed process}

In this section, we focus on comparing with the existing process, whether our two designed components could \textcolor{mycolor}{promote \textit{user engagement level}, reduce \textit{remediation duration}, and improve \textit{user satisfaction}}.

\subsubsection{\textcolor{mycolor}{User engagement level promotion}}

\begin{figure*}[ht]
	\centering
		\includegraphics[scale=.3]{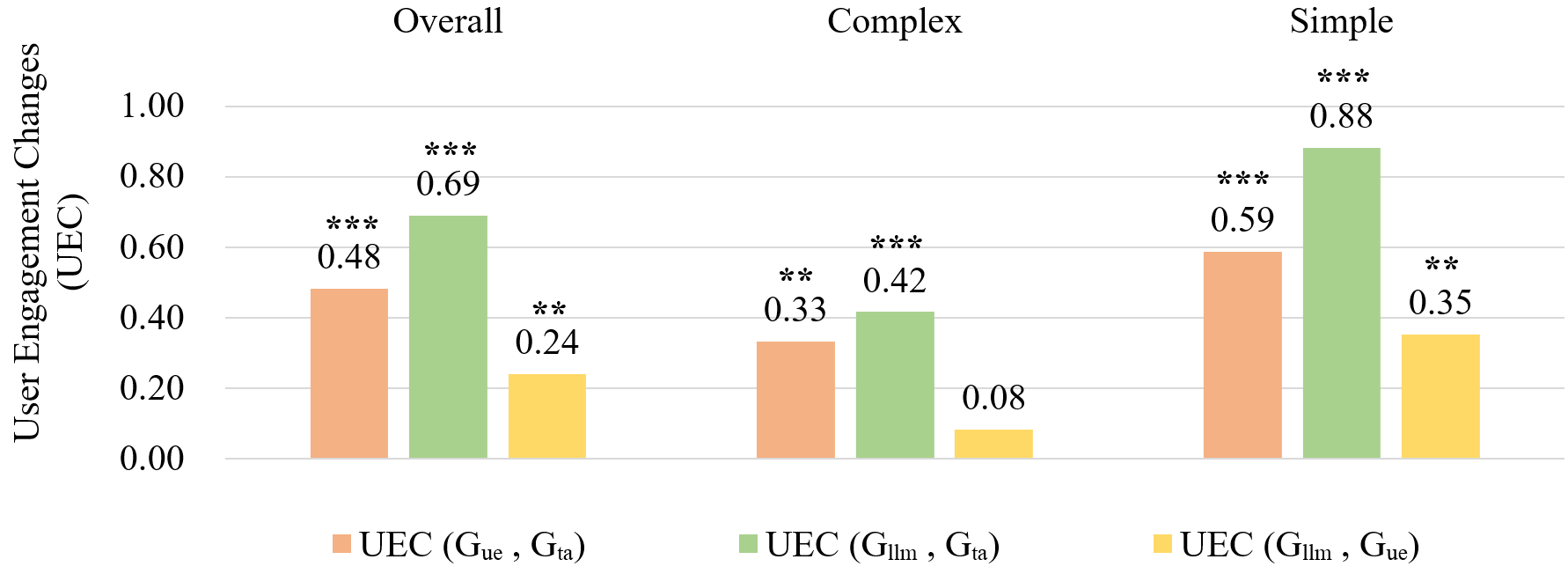}
	\caption{User engagement level changes between any two conditions and within different vulnerability groups. $*p<0.1,**p<0.05,***p<0.01.$}
	\label{FIG:vulengagement}
\end{figure*}

As reported in the left part of Figure \ref{FIG:vulengagement}, both $G_{ue}$ and $G_{llm}$ significantly improve the engagement with users, as they both demonstrate a significant positive user engagement level change when comparing with $G_{ta}$ ($p=0.000, d=1.342$, and $p=0.000, d=2.072$ respectively). Interestingly, we observe a significant positive user engagement level change between $G_{llm}$ and $G_{ue}$ ($p=0.045
, d=0.462$), meaning the introduction of LLM for security technicians can empower users' engagement level when handling vulnerabilities collaboratively. This surprise observation indicates LLM's ability to provide more user-friendly vulnerability remediation solutions, although refined by technicians, which encourage users to take a more active role in vulnerability remediation.

More specifically, as shown in the right part of Figure \ref{FIG:vulengagement}, we observe the same positive effect for engaging users when handling \textit{complex} and \textit{simply} vulnerabilities. However, comparing $G_{llm}$ with $G_{ue}$, we found a significant positive user engagement level change  ($p=0.027, d=0.834$) when handling \textit{simple} vulnerabilities. The user engagement level change  ($p=0.361, d=0.198$) when handling \textit{complex} vulnerabilities is not significant but still larger than 0. In fact, our data shows that comparing 
$G_{llm}$ with $G_{ue}$, more simple vulnerabilities are remediated by users with limited security technicians' support, while more complex vulnerabilities are handled through the \textit{Technicians \& Users} collaborative model. One potential reason is that the remediation solutions for simple vulnerabilities can be more straightforward and easier to follow so that users can handle them without too much support from security technicians. Hence, the benefit of introducing LLM for improving users' engagement is more effective for \textit{simple} vulnerabilities like SSL/SSH vulnerabilities, but less for the \textit{complex} vulnerabilities like Mysql/Apache/Rare vulnerabilities.

\textbf{In summary, \textit{user engagement enhancement} and \textit{LLM-supported technician enhancement} can effectively improve user engagement levels in the vulnerability remediation process. While the additional LLM support can further empower the engagement with users, such an effect is more positive for patching simple vulnerabilities like SSL / SSH.}

\subsubsection{\textcolor{mycolor}{Remediation duration reduction}}

\begin{figure*}[ht]
	\centering
		\includegraphics[scale=.3]{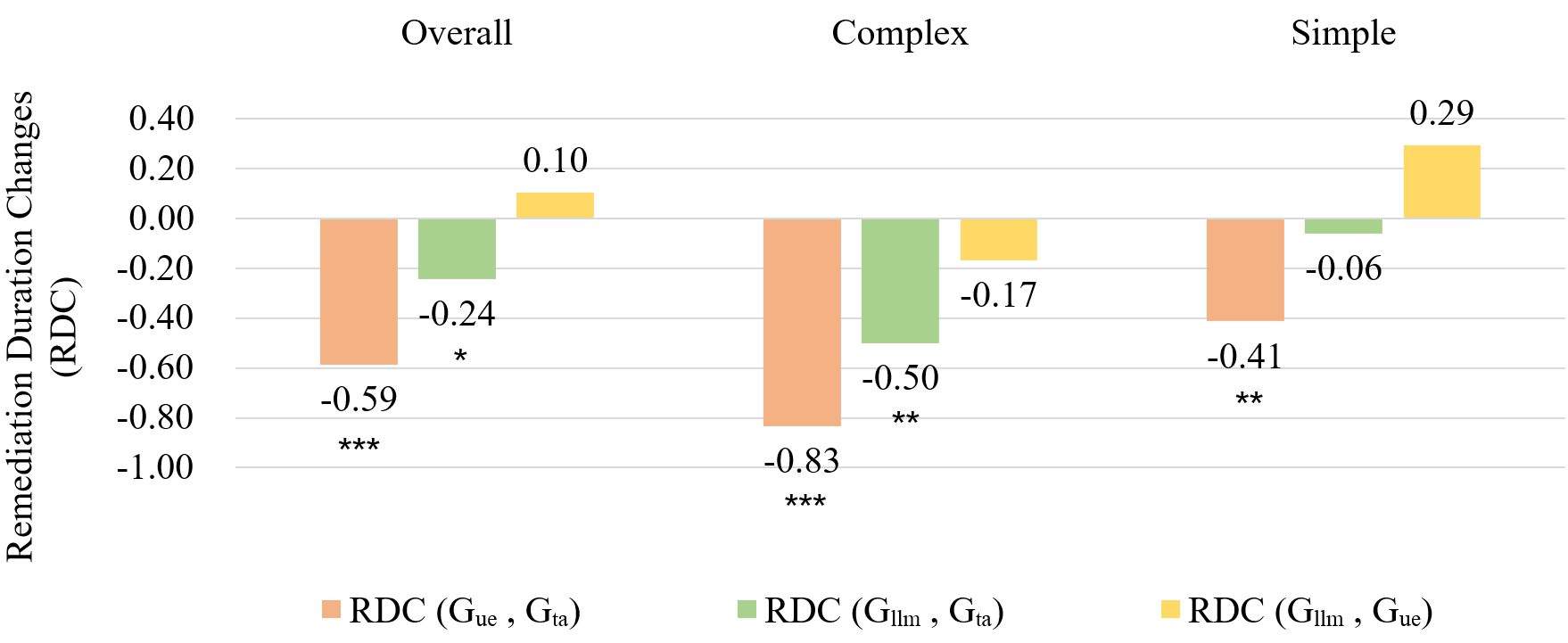}
	\caption{Remediation duration changes between any two conditions. "Overall" represents the whole 29 vulnerabilities, while others represent the two groups according to the vulnerability remediation complexity. \textit{*p<0.1,**p<0.05,***p<0.01.}
 }
	\label{FIG:Category}
\end{figure*}

As shown in the left part in Figure \ref{FIG:Category}, when comparing with the existing vulnerability remediation process $G_{ta}$, the remediation duration changes for $G_{ue}$ and $G_{llm}$ are both significantly lower than 0 ({$p=0.000, d=1.005$} and {$p=0.099, d=0.346$}).  This confirms that our optimized process can significantly reduce the time for vulnerability remediation. However, we observe an increasing, but not significant, remediation duration change when comparing $G_{llm}$ and $G_{ue}$. This suggests that providing additional LLM support for technicians may require security technicians to adapt to LLM, which needs additional time spending. 

We then investigate how the vulnerability complexity impacts the reduction of remediation duration. As shown in the right part of Figure \ref{FIG:Category}, we observe a consistent remediation duration reduction when comparing $G_{ue}$ and $G_{llm}$ with $G_{ta}$. More interestingly, for \textit{complex vulnerabilities}, such a reduction is more significant. We also observe a negative, though not significant, $RDC(G_{llm}, G_{ue})$. This indicates that additional LLM support can further reduce, though not significant in our study, the remediation duration when fixing \textit{complex vulnerabilities}. In contrast, for \textit{simple} vulnerability, the benefit from user engagement and additional LLM-supported technician enhancement decline, where $RDC(G_{llm},G_{ta})$ even become insignificant and $RDC(G_{llm},G_{ue})$ demonstrates a positive value. This suggests that the time consumption from adopting LLM by security technicians surplus its potential benefit when handling \textit{simple} vulnerabilities.

\textbf{\textcolor{mycolor}{In summary, \textit{user engagement enhancement} and \textit{LLM-supported technician enhancement} can significantly reduce the remediation duration.} However, while the additional LLM support for security technicians is not sufficient in reducing remediation duration, especially for \textit{simple} vulnerabilities like SSL/SSH, we observe its transformative potential in handling \textit{complex} vulnerabilities like Mysql/Apache/Rare vulnerabilities.}

\subsubsection{\textcolor{mycolor}{User satisfaction improvement}}

In Study 2, 87 users completed the survey, and we followed the same procedure to remove the invalid responses, resulting in 85 responses (28 from $G_{ta}$, 28 from $G_{ue}$, and 29 from $G_{llm}$) for analysis. 

\begin{figure*}[ht]
	\centering
		\includegraphics[scale=.3]{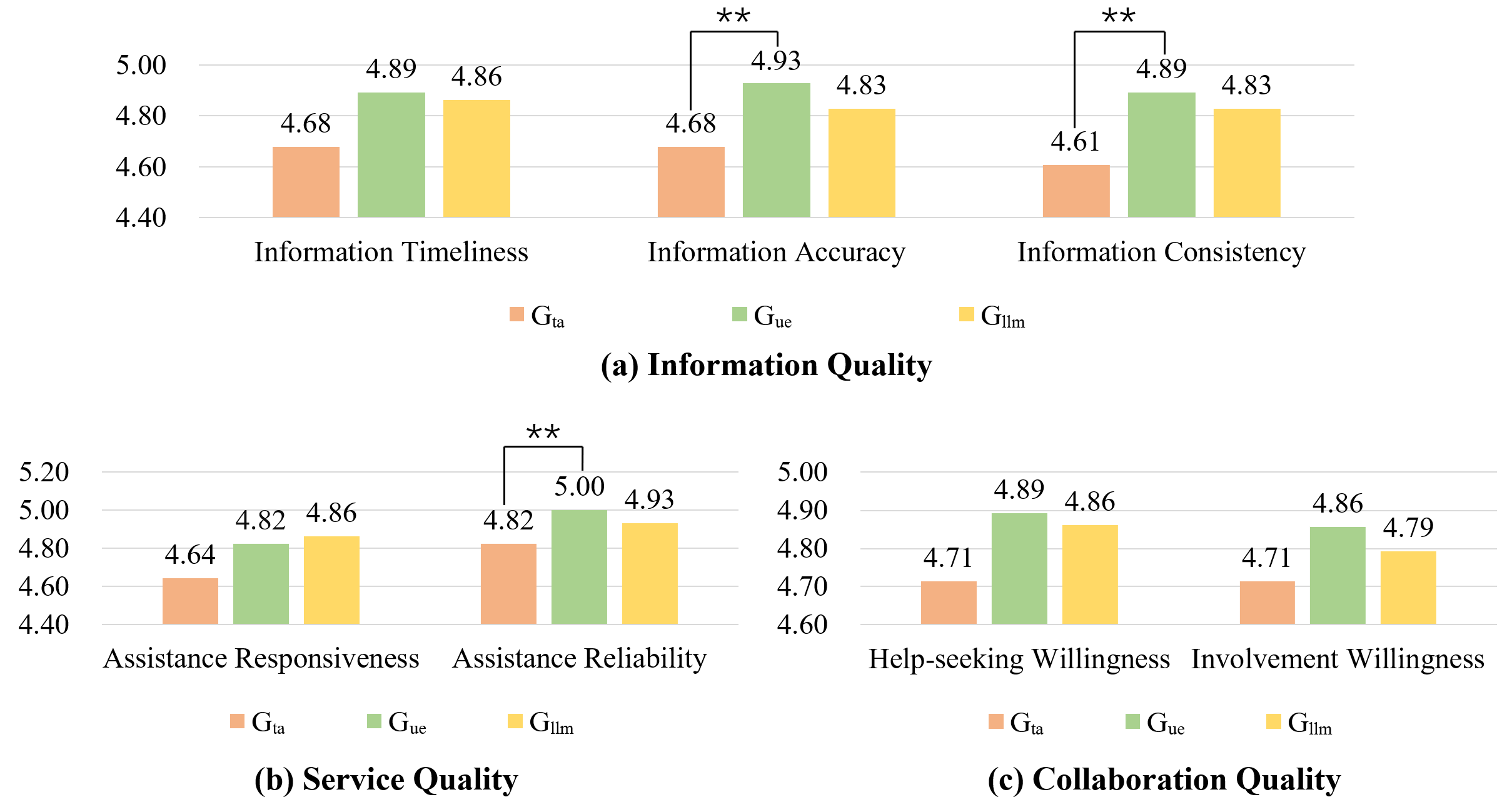}
	\caption{User satisfaction improvement regarding information, service, and collaboration quality. \textit{*p<0.1,**p<0.05,***p<0.01.}}
	\label{FIG:study2result}
\end{figure*}

As reported in Figure \ref{FIG:study2result}, we observe a higher value in all metrics from $G_{ue}$ and $G_{llm}$ when compared with $G_{ta}$, indicating that overall our designed process can improve the user satisfaction. More specifically, users from condition $G_{ue}$ rated the information accuracy (\textit{p=0.03,d=0.647}) and consistency (\textit{p=0.04,d=0.623}) significantly higher than $G_{ta}$. Compared to $G_{ta}$, $G_{ue}$ demonstrated significantly higher assistance reliability (\textit{p=0.032,d=0.651}), meaning participants report that the collaboration with security technicians could address vulnerability remediation issues more comprehensively. 

However, when comparing $G_{ue}$ and $G_{llm}$, there exist no significant differences, meaning that providing additional LLM support for security technicians in developing vulnerability plans did not make a significant change on user satisfaction in our study. This is reasonable as we did not provide LLM for users and did not tell users that the remediation solution is developed with LLM support. We even observe a decline, although not significant, for all metrics except assistance responsiveness, suggesting that an in-depth investigation of the LLM's roles in this collaborative vulnerability remediation process can be helpful, which we will report in Study 3.

\textbf{In summary, \textit{user engagement enhancement} can improve user satisfaction, especially in information accuracy, consistency, and assistance reliability. However, the additional LLM support for technicians has a limited, if not negative, impact on user satisfaction.}

\section{Study3: LLM's roles in collaborative vulnerability remediation process}

Our observations in Study 2 demonstrate the undoubted effectiveness of user engagement enhancement, \textcolor{mycolor}{confirming our expectation.} \textcolor{mycolor}{However, the additional LLM support exhibits complex heterogeneous impacts. 
Hence, we conduct a follow-up study to delve deep into how LLMs influence the collaboration between security technicians and users, as well as the subsequent effects of LLMs on the efficiency of the vulnerability remediation process, aiming at providing hints for future process optimization.} More specifically, we conduct an analysis on the 29 remediation reports submitted by security technicians from condition $G_{llm}$, which provides LLMs to security technicians, and semi-structural interviews with 12 users and 6 technicians participating in our Study 2 from all three conditions.

\subsection{Dataset and analysis method}

\subsubsection{Remediation report}

For condition $G_{llm}$ in Study 2, \textcolor{mycolor}{the remediation report submitted by security technicians includes} how the solutions provided by LLMs are adopted, the technicians' assessment of their effectiveness, as well as their experiences in using LLMs to support the vulnerability remediation. More specifically, out of the 29 remediation reports, 93.1\% (=26/29) chose to adopt the LLM-suggested solutions, with 58.6\%(=17/29) rating these solutions as effective at 70\% or higher and 9 reports (31\%) considered the LLM solutions to be effective in the range of 30\% to 70\%. Three tasks (10.3\%) believed the LLM-provided solutions were effective at 30\% or below. This diverse rating and their experiences provide us with details to examine LLM's roles, strengths, weaknesses, and challenges in supporting vulnerability remediation tasks.

\subsubsection{Semi-structured interview}

We also conducted a semi-structured interview to explore the collaborative dynamics, especially the interaction with LLMs, and the role of LLMs in facilitating collaboration among security technicians and users. We invited 18 participants, consisting of 6 security technicians and 12 users, where each condition ($G_{ta}$, $G_{ue}$, and $G_{llm}$) is represented by 2 technicians and 4 users. In particular, our interviews included open-ended questions focusing on how the security technicians, users, and LLM collaborate with each other, the perceived strengths and weaknesses of LLM's capability in supporting vulnerability remediation, the challenges to adopting the LLM tools, and the interactions with LLMs. At the start of each interview, participants will be provided with informed consent forms. We also told them that all questions were optional and that their anonymity and privacy would be ensured. 

\subsubsection{Data analysis}

Following the previous studies (\cite{kuvcera2021bedtime,nicholson2019if,arrambide2022don,alfrink2023contestable,van2023towards}), we adopted the thematic analysis method to analyze our interview data and feedback from the submitted remediation reports. Two independent researchers analyzed all the data thoroughly to ensure unbiased results. Then, the group discussions with an additional researcher were set up to thoroughly and collaboratively review, edit, and come to an agreement on the extracted themes. 

\textcolor{mycolor}{\subsection{LLMs' double-edged role in the collaborative vulnerability remediation process}}

\textcolor{mycolor2}{In this section, we will discuss the role of LLMs on user engagement, remediation duration, and user satisfaction in the vulnerability remediation process by comparing the conclusions drawn from $G_{llm}$ and $G_{ue}$. However, we acknowledge that other potential factors besides LLMs might influence these measurements.}

\textcolor{mycolor2}{For instance, while the quality of the remediation plan generated by LLMs is a significant factor affecting user engagement, other elements, such as users' technical knowledge and overall confidence, may also play a crucial role. However, we conducted the experiment with controlled conditions to limit the impacts of these factors. While all groups handle the same sets of vulnerabilities, relevant users are also randomly assigned to different groups, ensuring no significant differences among groups. }

\textcolor{mycolor2}{Furthermore, LLMs influenced the collaboration models between technicians and users, leading to the increasing user involvement in the vulnerability remediation process, and the differences in experience and expertise between technicians and users can affect the final remediation duration. 
However, the increasing user engagement can considered a result of LLMs-generated vulnerability remediation reports, which include user-friendly and step-by-step guidance, code snippets, and solutions. Therefore, the impact of LLMs on remediation duration includes both the direct effects of LLM support and the indirect effects through increasing user engagement.}

\textcolor{mycolor2}{Overall, we can conclude that the primary distinction between $G_{llm}$ and $G_{ue}$ lies in whether LLM support is provided. Consequently, in the subsequent subsections, we will focus on the role of LLMs in collaborative vulnerability remediation.}

\subsubsection{\textcolor{mycolor}{Bright sides of LLMs}}

\textcolor{mycolor}{The findings from Study2 suggest that the additional LLM support could significantly improve the users' participation, especially for simple vulnerabilities, and slightly reduce the remediation duration for complex vulnerabilities. To explore the underlying reasons for the above two conclusions in Study2, we conducted a detailed review of remediation reports and semi-structured interview transcripts.}

\textcolor{mycolor}{\textit{\textbf{(1) Why could LLMs significantly enhance user engagement, particularly regarding simple vulnerabilities?}}}

\textcolor{mycolor}{The primary reason is that they can \textbf{provide user-friendly and step-by-step guidance, code snippets, and solutions.} Firstly, for security technicians, the clarity and richness of information provided by the LLM tool made it an excellent reference. This strength, especially for vulnerabilities with mature remediation solutions, has been consistently mentioned in many reports (i.e., \textit{T2, T3, T9, T10, T12, T16, T18, T21, T26 \& T27}):}

\begin{quote}
    \textit{"LLMs consistently offer clear guidance, key code segments, and detailed step-by-step instructions, particularly for mature patches"}
\end{quote}

In some cases, LLMs can provide details such as the impacted version from a given vulnerability, which is critical in vulnerability remediation tasks. This will significantly improve the usability of the generated solution and lead to successful remediation outcomes. For example, the remediation report \textit{T25} highlights that:

\begin{quote}
    \textit{"LLMs prove effective in providing vulnerability's impact version information"}
\end{quote}

\textcolor{mycolor}{More importantly, security technicians can now provide user-friendly vulnerability remediation solutions with greater ease, enabling users to follow the solutions step-by-step to remediate the vulnerabilities.} For example, users \textit{U10} and \textit{U11} who are both from $G_{llm}$ mentioned that:

\begin{quote}
    \textit{"The security technicians provide us the remediation solution. As suggested by the technicians, we follow the solution step-by-step, and it succeeds [...] although we may be slower than the technician did it by themselves."}
\end{quote}

% 虽然 LLM 生成的修复计划的质量是一个重要因素，但其他因素（例如用户的技术知识和整体信心）也可能发挥关键作用。因此，重要的是要考虑到用户参与度可能受到各种因素的影响。

% 我们承认，在Section5.2.1关于用户参与的讨论中，没有给出所有可能影响用户参与程度的因素。虽然 LLM 生成的修复计划的质量是影响用户参与程度的一个重要因素，但其他因素（例如用户的技术知识和整体信心）也可能发挥关键作用。

% 但是我们用控制变量的方式来开展实验，即对于用户的技术知识和整体信心、修复漏洞的难易程度、修复设备的类型等外部变量，均采用随机分配或者保持相同的方式，使得三个实验组的这些外部变量没有明显区别。

% Section5.2主要解释了G_llm与G_ue相比得到的结论，在外部变量保持一致的条件下，G_llm与G_ue相比，最大的区别就是G_llm有LLM的支持。因此，在Section5.2中，我们会重点强调LLM在漏洞修复流程中起到的作用，我们在Section5.2.1声明了这一点。 

% \textcolor{mycolor2}{Note that while the quality of the remediation plan generated by the LLM is a significant factor influencing the level of user engagement, other elements, such as users' technical knowledge and overall confidence, may also play a crucial role. However, we conducted the experiment with controlled conditions. We randomized or kept constant external variables such as users' technical knowledge and overall confidence, the complexity level of vulnerability, and the types of remediation devices, ensuring no significant differences among the three experimental groups. Under consistent external conditions, the main difference between $G_{llm}$ and $G_{ue}$ is the support provided by LLMs. Therefore, in this section, we emphasized discussing the role of LLMs in facilitating user engagement, with subsequent sections following a similar approach.} 

\textcolor{mycolor}{\textit{\textbf{(2) Why could LLMs slightly reduce the remediation duration for complex vulnerabilities?}}}

\textcolor{mycolor}{One of the reasons is that LLMs could \textbf{expediently and efficiently generate preliminary, good-quality solutions as the starting point. }Security technicians emphasize the cumbersome associated with existing vulnerability remediation work, which often requires time-consuming searches on various IT forums for supplementary information or practical case studies alongside downloading official patches. In contrast, LLM tools can generate a preliminary solution consolidating related resources, which may be scattered throughout different sources, especially for those complex vulnerabilities. This will reduce the time consumed by security technicians when developing the remediation solution. As elaborated in \textit{T1} related to Apache vulnerability:}

\begin{quote}
    \textit{"LLMs can swiftly provide relatively complete solutions for vulnerabilities, eliminating the need for extensive resource hunting"}
\end{quote}

\textcolor{mycolor}{Another reason is that LLMs can \textbf{bring forth novel perspectives and surface previously overlooked issues.} Security technicians acknowledge that LLMs can offer multiple solutions for the same remediation tasks. Among these solution candidates, there are some novel approaches that are rarely encountered, which can be combined to generate a more comprehensive solution. This can expedite the vulnerability remediation process, especially for those complex vulnerabilities. For instance, the remediation report \textit{T7} related to Apache vulnerability mentioned:}

\begin{quote}
    \textit{"LLMs suggest a complete upgrade as one solution, and a less common but new approach by modifying source code without upgrade"}.
\end{quote}

\subsubsection{\textcolor{mycolor}{Dark sides of LLMs}}

\textcolor{mycolor}{As shown in Study2, the additional LLM support increases the remediation duration for simple vulnerabilities and has a limited, if not negative, impact on users' satisfaction. Likewise, we meticulously examined remediation reports and interview data, striving to discern the underlying logic for the aforementioned observations.}

\textcolor{mycolor}{\textit{\textbf{(1) Why do LLMs increase the remediation duration for simple vulnerabilities?}}}

\textcolor{mycolor}{Firstly, the solutions provided by LLMs can be \textbf{too general and without considering personalized environment compatibility}, resulting in more time for remediation. While LLMs can provide general solutions, they may not adequately account for the unique configurations and constraints of specific environments. This has been expressed in several reports related to those simple (SSL / SSH) vulnerabilities (i.e. \textit{T17 \&T21}):}

\begin{quote}
    \textit{"The solutions offered by LLMs for similar vulnerabilities tend to be generic and may not adequately anticipate the unique risks associated with individualized environments. [...] it needs further customization and human intervention to avoid unintended consequences on application functionality. "}
\end{quote}

\textcolor{mycolor}{Secondly, solutions generated by LLMs sometimes tended to be \textbf{overly macroscopic and idealistic}. These solutions, while conceptually sound, were challenging to implement in real-world security contexts due to missing information for quality verification, requiring unavailable resources, or being too time-consuming. For example, \textit{T27} related to SSL vulnerability mentioned:}

\begin{quote}
    \textit{"Certain solutions provided by LLMs presented by the tool incurred high time costs and were less feasible in real-world scenarios"}
\end{quote}

\textcolor{mycolor}{Thirdly, the solutions provided by LLM tools were \textbf{not always complete or highly executable}. Security experts found that further refinement and validation were often necessary to transform these solutions into practical, implementable solutions. For instance, \textit{T16} related to SSL vulnerability mentioned:}

\begin{quote}
      \textit{"Certain solutions lacked algorithm details, necessitating manual assessment after running 'openssl ciphers –v'"} 
\end{quote}

\textcolor{mycolor}{These observations highlight that LLMs should be a complementary tool to human expertise rather than a standalone solution. While LLM's insights are valuable, security technicians must be equipped with critical judgment capability and expertise to bridge the gap between idealized suggestions and the practical realities of their specific environments. Additional tools, such as testing environments to validate the generated solutions and effective collaborative channels to communicate with users, become critical complementary components to empower LLM-support tools for vulnerability remediation.}

\textcolor{mycolor}{Last but not least, for remediation of simple vulnerabilities, while the user involvement is higher, as newcomers to cybersecurity, these users need to learn the relevant knowledge and techniques to follow the remediation plan. In other words, the user's remediation process would be slower than that of security technicians. Additionally, some users noted that the time required for communication with the security technicians could increase the overall remediation duration. For example, users \textit{U11} and \textit{U12} from $G_{llm}$ mentioned that: }

\begin{quote}
      \textcolor{mycolor}{\textbf{[U11]:}} \textit{\textcolor{mycolor}{"This is my first time patching a vulnerability, and with the technicians guiding me throughout the process.[...], my remediation is certainly slower than those of technicians."}} 
      
      \textcolor{mycolor}{\textbf{[U12]:}} \textit{\textcolor{mycolor}{"[...], and communicating with technicians incurs a time cost."}} 
\end{quote}

\textcolor{mycolor}{However, it is worth noting that some users with foundational knowledge in cybersecurity feel that upon receiving alert information, being provided with remediation documentation to fix the vulnerability independently, has increased the efficiency of the remediation process. For instance, user \textit{U10} from $G_{llm}$ mentioned that: 
}
\begin{quote}
      \textit{\textcolor{mycolor}{"Previously, upon receiving vulnerability alerts, I wasn't provided with a corresponding remediation plan. However, this time, armed with a remediation document from the technicians, I successfully resolved the issue upon review and execution, which I believe has enhanced the efficiency of the remediation process. "}} 
\end{quote}

\textcolor{mycolor}{This comparison illustrates the critical importance of equipping users within an organization with fundamental knowledge and skills in cybersecurity. Compared to newcomers in the field, users with a baseline understanding of cybersecurity are more inclined to participate in the vulnerability remediation process and are able to contribute to enhancing the efficiency of the process.}

\textcolor{mycolor}{\textit{\textbf{(2) Why do LLMs have a limited, if not negative, impact on users' satisfaction?}}}

\textcolor{mycolor}{The primary reason is that while LLMs have effectively enhanced user engagement, some users believe that vulnerability remediation is not part of their job responsibilities, and involving them leads to decreased satisfaction with the collaborative vulnerability remediation process. In other words, the decline in user satisfaction is due to the negative impact of the immature organizational cybersecurity culture, rather than the effectiveness of the LLMs. For example, user \textit{U9} from $G_{llm}$ mentioned that:}

\begin{quote}
      \textit{\textcolor{mycolor}{"Our job involves running data through work software, but we're not experts in security work; it's better to leave the full responsibility of vulnerability remediation to dedicated security technicians."}} 
\end{quote}

\textcolor{mycolor}{However, while collaborating with security technicians to remediate vulnerabilities might indeed affect the satisfaction levels of some users, we do observe a considerable number of users indicating a willingness to engage in the vulnerability remediation workflow and are eager to acquire knowledge and improve their related skills. For example, user \textit{U11} from $G_{llm}$ mentioned that:}

\begin{quote}
      \textit{\textcolor{mycolor}{"It's great that security technicians are guiding us in vulnerability remediation; it's usually hard to learn these skills without someone to show us the ropes."}} 
\end{quote}

\textcolor{mycolor}{Moreover, in the existing process, security technicians are the main force supporting the collaborative vulnerability remediation process among themselves, users, and LLMs. If users can remediate vulnerabilities themselves or assist in the process, it would undoubtedly reduce the technicians' workload and improve the effectiveness of the process. Security technicians \textit{S5} and \textit{S6} from $G_{llm}$ both mentioned this aspect:}

\begin{quote}
    \textcolor{mycolor}{\textbf{[S5]:}} \textit{\textcolor{mycolor}{"Ideally, we aim to teach users how to fix vulnerabilities themselves, seeking our assistance only for more complex vulnerabilities."}} 

    \textcolor{mycolor}{\textbf{[S6]:}} \textit{\textcolor{mycolor}{"It's much better if users can collaborate with us to fix vulnerabilities rather than us doing it alone. This is because we don't have extensive knowledge of their specific devices. Users can help us determine which solutions have the least impact on their systems, saving us much time."}} 
\end{quote}

\section{Discussions, implementations and limitations}

\subsection{\textcolor{mycolor}{Summary of findings}}

\begin{figure*}[ht]
	\centering
		\includegraphics[scale=.4]{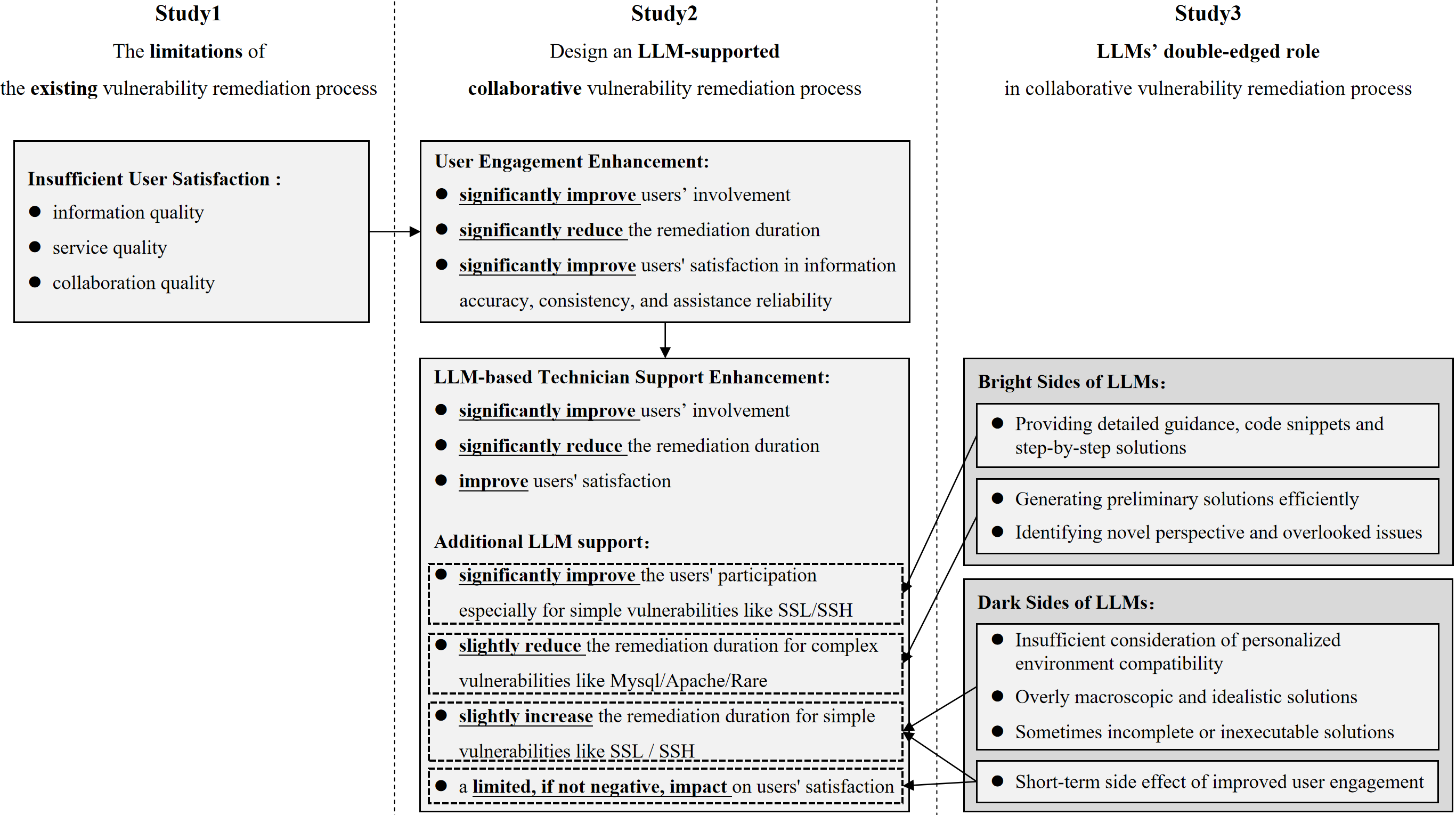}
	\caption{\textcolor{mycolor}{Summary of findings.}}
	\label{FIG:findings}
\end{figure*}

\textcolor{mycolor}{As consolidated in Figure \ref{FIG:findings}, our three-step mixed-method study yielded several significant findings that correspond to our research questions. }

\textcolor{mycolor}{Firstly, to answer RQ1, building upon the D\&M model \citep{delone1992information,delone2003delone}, the survey in Study 1 uncovered insufficient user satisfaction regarding information, service, and collaboration quality within the existing vulnerability remediation process in practice.
Hence, we design the collaborative process among security technicians, users, and LLMs (Study 2), which enhances the existing process with \textit{user engagement enhancement} using alert information and remediation solution sharing, and \textit{LLM-supported technician enhancement} which provide LLM for remediation solution development. }

\textcolor{mycolor}{Secondly, to answer RQ2, we conduct a field study to validate the effectiveness of the designed collaborative vulnerability remediation process among security technicians, users, and LLMs. In practice, it could effectively improve the existing process, including significantly promoting users' engagement, significantly reducing remediation duration, and improving user satisfaction. }

\textcolor{mycolor}{More specifically, to answer RQ2.1, \textit{user engagement enhancement} demonstrated the expected effectiveness from user involvement. 
Engaging users in the collaborative vulnerability remediation process can significantly improve user engagement levels, reduce the remediation duration, and improve user satisfaction, especially in information accuracy, consistency, and assistance reliability.}

\textcolor{mycolor}{
Additionally, to answer RQ2.2, the additional LLM support significantly improves the user engagement level, especially for simple vulnerabilities like SSL/SSH. In other words, LLMs could undoubtedly encourage more users to engage in the vulnerability remediation process. As for the impact on existing workflow, LLMs demonstrate the potential to reduce remediation duration for complex vulnerabilities but show limited or even negative effects on users' satisfaction improvement.}

\textcolor{mycolor}{
Finally, to answer RQ3, the thematic analysis of the remediation reports and the follow-up semi-structured interviews (Study 3) uncover the reason behind LLMs' double-edged role in the optimized process. On the bright side, LLMs could significantly enhance user engagement due to their ability to offer user-friendly, step-by-step guidance, code snippets, and solutions. LLMs could also further slightly reduce the remediation duration for complex vulnerabilities due to their capability to efficiently generate high-quality preliminary solutions, as well as their ability to introduce fresh perspectives and surface previously overlooked issues.}

\textcolor{mycolor}{
On the dark side, LLMs increase the remediation duration for simple vulnerabilities because they may insufficiently consider personalized environment compatibility and occasionally offer overly macroscopic, idealistic, or inexecutable solutions, which require significant extra efforts from security technicians to refine and redevelop the solutions generated by LLMs. Additionally, as users are not as professional as security technicians, their involvement in the remediation process would increase the time needed to communicate and adopt the remediation plan. What's more, some users perceive it as outside their job scope. In other words, the insignificant effect of additional LLM support in reducing remediation duration for simple vulnerabilities and improving user satisfaction can be due to the side effect of its improvement in improving user engagement. Yet, as we observe many users' willingness to participate and learn, along with enabling security technicians to support multi-users simultaneously, we would expect that such an insufficient effect could be just a short-term hiccup.}

\textcolor{mycolor}{
Overall, we contribute a practically validated LLM-supported collaborative process that improves the effectiveness of vulnerability remediation, a real-world high-stakes task in practice. Additionally, we reveal the underlying mechanisms of LLMs' double-edged role in the refined process, providing valuable insights for enhancing security practices.}

\subsection{Lessons learn from implementing the LLM-supported collaborative process}

Beyond designing the LLM-supported collaborative process to improve the effectiveness of vulnerability remediation, our findings highlight the below lessons that we could follow when incorporating LLM for other high-stakes tasks.

\subsubsection{Making it a top priority to facilitate collaborations among all associated stakeholders, including the LLM} As shown in our study, the \textit{user engagement enhancement} component contributes the most significant improvement in the optimized process. \textcolor{mycolor}{While the role of additional LLM support is a double-edged sword, LLMs can undoubtedly motivate users' involvement in the vulnerability remediation process and promote overall user engagement, as they are capable of generating user-friendly and step-by-step solutions.} More importantly, even for LLMs themselves, when integrated into practical processes like vulnerability remediation, they need to collaborate with complementary tools such as solution quality evaluations and expertise for iterative exploration. Hence, to enhance the effectiveness of high-stakes tasks in practice, facilitating collaborations among all associated stakeholders, including security technicians, users, and the LLM, should be the top priority.

\subsubsection{\textcolor{mycolor}{LLM's performance can be influenced by task complexity, so reshaping the role of LLM according to task levels could be helpful}}  Our study demonstrates the unequal effects of LLM's support for simple and complex vulnerability remediation. LLM is more effective in reducing the remediation duration for complex vulnerabilities but may increase it for simple vulnerabilities. \textcolor{mycolor}{The probable cause is that for simple vulnerabilities, security technicians, leveraging their rich experience, can already offer remediation solutions swiftly, making it challenging for LLMs to further reduce the remediation duration. Additionally, the limitations of LLMs in generating overly idealistic, incomplete, and misleading information can be amplified. This forces technicians to spend extra time verifying the feasibility of LLM-generated remediation plans, ultimately increasing the overall vulnerability remediation time. Hence, the capacity for LLMs to augment the efficiency of security experts in addressing simple vulnerabilities is quite limited. However, leveraging LLMs to support users—novices in cybersecurity— to independently remediate simple vulnerabilities could be a promising avenue for organizations to enhance cybersecurity process efficiency as LLMs can generate more user-friendly remediation plans for simple vulnerabilities.}

\textcolor{mycolor}{Conversely, for complex vulnerabilities, the remediation process is inherently lengthy, and security technicians may not have hands-on solutions. Hence, LLMs can quickly generate preliminary remediation plans, significantly reducing the time security experts spend on gathering information and ultimately shortening the overall remediation duration. This lays a solid foundation for future research on how LLMs can better collaborate with security experts to further optimize the remediation process for complex vulnerabilities.}

\textcolor{mycolor}{In summary, the role of LLMs in practices should adapt to the complexity of the tasks.} This is aligned with the existing study investigating the LLM's application in customer services, which also found LLM' unequal support for employees with different expertise \citep{brynjolfsson2023generative}. Considering the limited availability of professional experts in practice, one promising solution is to support novice technicians with LLMs in identifying and polishing existing solutions for simpler tasks while tasking the experts with LLMs in solution development and validation, complex task solving, and interaction strategy development. However, how to support such a role rotation is still an open question. 

\subsubsection{\textcolor{mycolor}{The side effect of improved user engagement could be a short-term hiccup, approaching it with a rational mindset}}

\textcolor{mycolor}{LLMs can assist in generating user-friendly and clear vulnerability remediation plans, encouraging more users to participate in the remediation process. However, the increased user involvement can result in side effects, such as a slight decrease in user satisfaction and longer remediation duration for simple vulnerabilities. On the one hand, some users feel that vulnerability remediation is an additional workload beyond their in-role responsibilities, leading to a decline in overall user satisfaction. However, it is crucial to make all users within the organization aware that maintaining the organization's cybersecurity is part of their in-role responsibilities, which is essential for fostering a strong cybersecurity culture \citep{huang2019technology,li2019investigating,posey2020exploratory}. On the other hand, some users who are willing to be involved in the remediation process will require more time and even require step-by-step guidance from technicians, reducing overall efficiency. Therefore, educating employees about cybersecurity, such as skills in following security guidance to fix vulnerabilities, can enhance the organization's overall cybersecurity capabilities \citep{he2020improving,kweon2021utility}.}

\textcolor{mycolor}{Overall, given LLM's capability to improve user engagement and its continuous improvement in generating remediation solutions, accompanied by organizations' efforts in cultivating the cybersecurity culture and training users with necessary cybersecurity skills, we would expect that the short-term side effects of increased user participation would be diminished and the long-term benefit would be boosted.} 

\subsection{Limitations and future directions}

Our current study has several limitations, which open venues for future studies. First, like other research based on field study \citep{alan2014field,kuvcera2021bedtime}, our findings are undoubtedly constrained by practical circumstances. Specifically, we involve only six security technicians, although having six technicians for vulnerability remediation is already considered a luxurious security team for many organizations. Additionally, we can not control the personal characteristics, such as users' gender and age, as well as security technicians' prior knowledge of some vulnerabilities \citep{toet2016effects}. Furthermore, performing tasks at different times may introduce biases due to fluctuations in individuals' cognitive abilities and energy levels \citep{lewandowska2018would}. While we ensure that the survey, remediation report, and follow-up interviews are anonymous and include the attention checks following the best practices, sample bias, and self-report bias may still exist, especially for those subjective responses. Finally, our study categorizes the 29 selected vulnerabilities into two groups based on their remediation complexity. However, even within the same category, the remediation complexity may differ slightly. Therefore, further studies to deploy our designed collaborative vulnerability remediation process in larger-scale organizations involving more users and security technicians in handling more vulnerabilities to generalize our findings would be valuable.

Second, we ran our three-step study within one month, including a one-week survey, a two-week field study, and a one-week follow-up interview. While this provides us with rich information to investigate the process, some carry-over effects may exist between any two steps, and the one-month study period may still be short. In fact, the security technicians may improve their capability to interact with LLMs to generate more user-friendly and effective solutions in the longer term, as we have already observed some emerging interaction patterns during our study period. Therefore, further studies on the longer-term effect would provide additional insightful knowledge.

Third, we chose the vulnerability remediation task, a typical high-stakes task, as our study context. While our findings have been successfully implemented in other LLM-supported cybersecurity operations within the institution, we should acknowledge that with our context, the final solution generated with LLM support for vulnerability remediation and other cybersecurity operations can be verified. However, for many other high-stakes tasks, the impact of the outcome may not be explicit and cannot be validated in the short term. This will complicate the collaboration with LLMs, and our conclusions may not be general enough. Hence, studies would be valuable in extending our findings to such settings. 

\textcolor{mycolor}{Fourth, our experiment did not include a condition with only LLM-supported enhancement available, as our study primarily focused on how LLMs could further facilitate collaboration between security technicians and users. Interestingly, we reveal the double-edged impact of additional LLM support, especially the side effects of involving users in the process. Beyond improving LLM's capability to develop remediation plans for security technicians, future studies could further explore how to improve users' willingness and capabilities to mitigate the side effects and achieve long-term benefits.}

Last but not least, in this study, we follow the industrial practices to group two security technicians. While we consider them as a whole to interact with LLMs, taking a closer look into how these two technicians work together to interact with LLMs could enrich our understanding of group decisions with LLMs' support \citep{chiang2023two}. Also, the LLM-based tool is only provided to security technicians to avoid overwhelming users. Further studies to provide optimized LLM tools for both users and security technicians to further empower their collaboration in the vulnerability remediation process would be a direction worth exploring.

\appendix

% \input{7.appendix}
% \section{My Appendix}
% Appendix sections are coded under \verb+\appendix+.

% \verb+\printcredits+ command is used after appendix sections to list 
% author credit taxonomy contribution roles tagged using \verb+\credit+ 
% in frontmatter.


\bibliographystyle{cas-model2-names}

\bibliography{cas-refs}

\begin{thebibliography}{118}
\expandafter\ifx\csname natexlab\endcsname\relax\def\natexlab#1{#1}\fi
\providecommand{\url}[1]{\texttt{#1}}
\providecommand{\href}[2]{#2}
\providecommand{\path}[1]{#1}
\providecommand{\DOIprefix}{doi:}
\providecommand{\ArXivprefix}{arXiv:}
\providecommand{\URLprefix}{URL: }
\providecommand{\Pubmedprefix}{pmid:}
\providecommand{\doi}[1]{\href{http://dx.doi.org/#1}{\path{#1}}}
\providecommand{\Pubmed}[1]{\href{pmid:#1}{\path{#1}}}
\providecommand{\bibinfo}[2]{#2}
\ifx\xfnm\relax \def\xfnm[#1]{\unskip,\space#1}\fi
%Type = Inproceedings
\bibitem[{Aghaei et~al.(2022)Aghaei, Niu, Shadid and Al-Shaer}]{aghaei2022securebert}
\bibinfo{author}{Aghaei, E.}, \bibinfo{author}{Niu, X.}, \bibinfo{author}{Shadid, W.}, \bibinfo{author}{Al-Shaer, E.}, \bibinfo{year}{2022}.
\newblock \bibinfo{title}{Securebert: A domain-specific language model for cybersecurity}, in: \bibinfo{booktitle}{International Conference on Security and Privacy in Communication Systems}, \bibinfo{organization}{Springer}. pp. \bibinfo{pages}{39--56}.
%Type = Article
\bibitem[{Akgun et~al.(2022)Akgun, Hizal and Cavusoglu}]{akgun2022new}
\bibinfo{author}{Akgun, D.}, \bibinfo{author}{Hizal, S.}, \bibinfo{author}{Cavusoglu, U.}, \bibinfo{year}{2022}.
\newblock \bibinfo{title}{A new ddos attacks intrusion detection model based on deep learning for cybersecurity}.
\newblock \bibinfo{journal}{Computers \& Security} \bibinfo{volume}{118}, \bibinfo{pages}{102748}.
%Type = Inproceedings
\bibitem[{Alahmadi et~al.(2022)Alahmadi, Axon and Martinovic}]{alahmadi202299}
\bibinfo{author}{Alahmadi, B.A.}, \bibinfo{author}{Axon, L.}, \bibinfo{author}{Martinovic, I.}, \bibinfo{year}{2022}.
\newblock \bibinfo{title}{99\% false positives: A qualitative study of $\{$SOC$\}$ analysts' perspectives on security alarms}, in: \bibinfo{booktitle}{31st USENIX Security Symposium (USENIX Security 22)}, pp. \bibinfo{pages}{2783--2800}.
%Type = Article
\bibitem[{Alan et~al.(2014)Alan, Costanza, Fischer, Ramchurn, Rodden and Jennings}]{alan2014field}
\bibinfo{author}{Alan, A.}, \bibinfo{author}{Costanza, E.}, \bibinfo{author}{Fischer, J.}, \bibinfo{author}{Ramchurn, S.}, \bibinfo{author}{Rodden, T.}, \bibinfo{author}{Jennings, N.R.}, \bibinfo{year}{2014}.
\newblock \bibinfo{title}{A field study of human-agent interaction for electricity tariff switching} .
%Type = Inproceedings
\bibitem[{Alfrink et~al.(2023)Alfrink, Keller, Doorn and Kortuem}]{alfrink2023contestable}
\bibinfo{author}{Alfrink, K.}, \bibinfo{author}{Keller, I.}, \bibinfo{author}{Doorn, N.}, \bibinfo{author}{Kortuem, G.}, \bibinfo{year}{2023}.
\newblock \bibinfo{title}{Contestable camera cars: A speculative design exploration of public ai that is open and responsive to dispute}, in: \bibinfo{booktitle}{Proceedings of the 2023 CHI Conference on Human Factors in Computing Systems}, pp. \bibinfo{pages}{1--16}.
%Type = Inproceedings
\bibitem[{Allen et~al.(1993)Allen, Ballman, Begg, Miller-Jacobs, Muller, Nielsen and Spool}]{allen1993user}
\bibinfo{author}{Allen, C.D.}, \bibinfo{author}{Ballman, D.}, \bibinfo{author}{Begg, V.}, \bibinfo{author}{Miller-Jacobs, H.H.}, \bibinfo{author}{Muller, M.}, \bibinfo{author}{Nielsen, J.}, \bibinfo{author}{Spool, J.}, \bibinfo{year}{1993}.
\newblock \bibinfo{title}{User involvement in the design process: why, when \& how?}, in: \bibinfo{booktitle}{Proceedings of the INTERACT'93 and CHI'93 Conference on Human Factors in Computing Systems}, pp. \bibinfo{pages}{251--254}.
%Type = Inproceedings
\bibitem[{Arrambide et~al.(2022)Arrambide, Yoon, MacArthur, Rogers, Luz and Nacke}]{arrambide2022don}
\bibinfo{author}{Arrambide, K.}, \bibinfo{author}{Yoon, J.}, \bibinfo{author}{MacArthur, C.}, \bibinfo{author}{Rogers, K.}, \bibinfo{author}{Luz, A.}, \bibinfo{author}{Nacke, L.E.}, \bibinfo{year}{2022}.
\newblock \bibinfo{title}{“i don’t want to shoot the android”: Players translate real-life moral intuitions to in-game decisions in detroit: Become human}, in: \bibinfo{booktitle}{Proceedings of the 2022 CHI Conference on Human Factors in Computing Systems}, pp. \bibinfo{pages}{1--15}.
%Type = Inproceedings
\bibitem[{Ashby et~al.(2023)Ashby, Webb, Knapp, Searle and Fulda}]{ashby2023personalized}
\bibinfo{author}{Ashby, T.}, \bibinfo{author}{Webb, B.K.}, \bibinfo{author}{Knapp, G.}, \bibinfo{author}{Searle, J.}, \bibinfo{author}{Fulda, N.}, \bibinfo{year}{2023}.
\newblock \bibinfo{title}{Personalized quest and dialogue generation in role-playing games: A knowledge graph-and language model-based approach}, in: \bibinfo{booktitle}{Proceedings of the 2023 CHI Conference on Human Factors in Computing Systems}, pp. \bibinfo{pages}{1--20}.
%Type = Inproceedings
\bibitem[{Bach and Twidale(2010)}]{bach2010involving}
\bibinfo{author}{Bach, P.M.}, \bibinfo{author}{Twidale, M.}, \bibinfo{year}{2010}.
\newblock \bibinfo{title}{Involving reflective users in design}, in: \bibinfo{booktitle}{Proceedings of the SIGCHI Conference on Human Factors in Computing Systems}, pp. \bibinfo{pages}{2037--2040}.
%Type = Article
\bibitem[{Bailey and Pearson(1983)}]{bailey1983development}
\bibinfo{author}{Bailey, J.E.}, \bibinfo{author}{Pearson, S.W.}, \bibinfo{year}{1983}.
\newblock \bibinfo{title}{Development of a tool for measuring and analyzing computer user satisfaction}.
\newblock \bibinfo{journal}{Management science} \bibinfo{volume}{29}, \bibinfo{pages}{530--545}.
%Type = Inproceedings
\bibitem[{Bansal et~al.(2019)Bansal, Nushi, Kamar, Lasecki, Weld and Horvitz}]{bansal2019beyond}
\bibinfo{author}{Bansal, G.}, \bibinfo{author}{Nushi, B.}, \bibinfo{author}{Kamar, E.}, \bibinfo{author}{Lasecki, W.S.}, \bibinfo{author}{Weld, D.S.}, \bibinfo{author}{Horvitz, E.}, \bibinfo{year}{2019}.
\newblock \bibinfo{title}{Beyond accuracy: The role of mental models in human-ai team performance}, in: \bibinfo{booktitle}{Proceedings of the AAAI Conference on Human Computation and Crowdsourcing}, pp. \bibinfo{pages}{2--11}.
%Type = Article
\bibitem[{Bommasani et~al.(2021)Bommasani, Hudson, Adeli, Altman, Arora, von Arx, Bernstein, Bohg, Bosselut, Brunskill et~al.}]{bommasani2021opportunities}
\bibinfo{author}{Bommasani, R.}, \bibinfo{author}{Hudson, D.A.}, \bibinfo{author}{Adeli, E.}, \bibinfo{author}{Altman, R.}, \bibinfo{author}{Arora, S.}, \bibinfo{author}{von Arx, S.}, \bibinfo{author}{Bernstein, M.S.}, \bibinfo{author}{Bohg, J.}, \bibinfo{author}{Bosselut, A.}, \bibinfo{author}{Brunskill, E.}, et~al., \bibinfo{year}{2021}.
\newblock \bibinfo{title}{On the opportunities and risks of foundation models}.
\newblock \bibinfo{journal}{arXiv preprint arXiv:2108.07258} .
%Type = Article
\bibitem[{van~den Broek et~al.(2019)van~den Broek, Sergeeva and Huysman}]{van2019hiring}
\bibinfo{author}{van~den Broek, E.}, \bibinfo{author}{Sergeeva, A.}, \bibinfo{author}{Huysman, M.}, \bibinfo{year}{2019}.
\newblock \bibinfo{title}{Hiring algorithms: An ethnography of fairness in practice} .
%Type = Techreport
\bibitem[{Brynjolfsson et~al.(2023)Brynjolfsson, Li and Raymond}]{brynjolfsson2023generative}
\bibinfo{author}{Brynjolfsson, E.}, \bibinfo{author}{Li, D.}, \bibinfo{author}{Raymond, L.R.}, \bibinfo{year}{2023}.
\newblock \bibinfo{title}{Generative AI at work}.
\newblock \bibinfo{type}{Technical Report}. National Bureau of Economic Research.
%Type = Article
\bibitem[{Cabitza et~al.(2017)Cabitza, Rasoini and Gensini}]{cabitza2017unintended}
\bibinfo{author}{Cabitza, F.}, \bibinfo{author}{Rasoini, R.}, \bibinfo{author}{Gensini, G.F.}, \bibinfo{year}{2017}.
\newblock \bibinfo{title}{Unintended consequences of machine learning in medicine}.
\newblock \bibinfo{journal}{Jama} \bibinfo{volume}{318}, \bibinfo{pages}{517--518}.
%Type = Inproceedings
\bibitem[{Cai et~al.(2019)Cai, Reif, Hegde, Hipp, Kim, Smilkov, Wattenberg, Viegas, Corrado, Stumpe et~al.}]{cai2019human}
\bibinfo{author}{Cai, C.J.}, \bibinfo{author}{Reif, E.}, \bibinfo{author}{Hegde, N.}, \bibinfo{author}{Hipp, J.}, \bibinfo{author}{Kim, B.}, \bibinfo{author}{Smilkov, D.}, \bibinfo{author}{Wattenberg, M.}, \bibinfo{author}{Viegas, F.}, \bibinfo{author}{Corrado, G.S.}, \bibinfo{author}{Stumpe, M.C.}, et~al., \bibinfo{year}{2019}.
\newblock \bibinfo{title}{Human-centered tools for coping with imperfect algorithms during medical decision-making}, in: \bibinfo{booktitle}{Proceedings of the 2019 chi conference on human factors in computing systems}, pp. \bibinfo{pages}{1--14}.
%Type = Inproceedings
\bibitem[{Chen et~al.(2023)Chen, Arunasalam and Celik}]{chen2023can}
\bibinfo{author}{Chen, Y.}, \bibinfo{author}{Arunasalam, A.}, \bibinfo{author}{Celik, Z.B.}, \bibinfo{year}{2023}.
\newblock \bibinfo{title}{Can large language models provide security \& privacy advice? measuring the ability of llms to refute misconceptions}, in: \bibinfo{booktitle}{Proceedings of the 39th Annual Computer Security Applications Conference}, pp. \bibinfo{pages}{366--378}.
%Type = Inproceedings
\bibitem[{Chiang et~al.(2023)Chiang, Lu, Li and Yin}]{chiang2023two}
\bibinfo{author}{Chiang, C.W.}, \bibinfo{author}{Lu, Z.}, \bibinfo{author}{Li, Z.}, \bibinfo{author}{Yin, M.}, \bibinfo{year}{2023}.
\newblock \bibinfo{title}{Are two heads better than one in ai-assisted decision making? comparing the behavior and performance of groups and individuals in human-ai collaborative recidivism risk assessment}, in: \bibinfo{booktitle}{Proceedings of the 2023 CHI Conference on Human Factors in Computing Systems}, pp. \bibinfo{pages}{1--18}.
%Type = Article
\bibitem[{Choe(1996)}]{choe1996relationships}
\bibinfo{author}{Choe, J.M.}, \bibinfo{year}{1996}.
\newblock \bibinfo{title}{The relationships among performance of accounting information systems, influence factors, and evolution level of information systems}.
\newblock \bibinfo{journal}{Journal of management information systems} \bibinfo{volume}{12}, \bibinfo{pages}{215--239}.
%Type = Article
\bibitem[{Cidral et~al.(2018)Cidral, Oliveira, Di~Felice and Aparicio}]{cidral2018learning}
\bibinfo{author}{Cidral, W.A.}, \bibinfo{author}{Oliveira, T.}, \bibinfo{author}{Di~Felice, M.}, \bibinfo{author}{Aparicio, M.}, \bibinfo{year}{2018}.
\newblock \bibinfo{title}{E-learning success determinants: Brazilian empirical study}.
\newblock \bibinfo{journal}{Computers \& education} \bibinfo{volume}{122}, \bibinfo{pages}{273--290}.
%Type = Article
\bibitem[{Curtis et~al.(1988)Curtis, Krasner and Iscoe}]{curtis1988field}
\bibinfo{author}{Curtis, B.}, \bibinfo{author}{Krasner, H.}, \bibinfo{author}{Iscoe, N.}, \bibinfo{year}{1988}.
\newblock \bibinfo{title}{A field study of the software design process for large systems}.
\newblock \bibinfo{journal}{Communications of the ACM} \bibinfo{volume}{31}, \bibinfo{pages}{1268--1287}.
%Type = Inproceedings
\bibitem[{De-Arteaga et~al.(2020)De-Arteaga, Fogliato and Chouldechova}]{de2020case}
\bibinfo{author}{De-Arteaga, M.}, \bibinfo{author}{Fogliato, R.}, \bibinfo{author}{Chouldechova, A.}, \bibinfo{year}{2020}.
\newblock \bibinfo{title}{A case for humans-in-the-loop: Decisions in the presence of erroneous algorithmic scores}, in: \bibinfo{booktitle}{Proceedings of the 2020 CHI Conference on Human Factors in Computing Systems}, pp. \bibinfo{pages}{1--12}.
%Type = Article
\bibitem[{DeLone and McLean(1992)}]{delone1992information}
\bibinfo{author}{DeLone, W.H.}, \bibinfo{author}{McLean, E.R.}, \bibinfo{year}{1992}.
\newblock \bibinfo{title}{Information systems success: The quest for the dependent variable}.
\newblock \bibinfo{journal}{Information systems research} \bibinfo{volume}{3}, \bibinfo{pages}{60--95}.
%Type = Article
\bibitem[{DeLone and McLean(2003)}]{delone2003delone}
\bibinfo{author}{DeLone, W.H.}, \bibinfo{author}{McLean, E.R.}, \bibinfo{year}{2003}.
\newblock \bibinfo{title}{The delone and mclean model of information systems success: a ten-year update}.
\newblock \bibinfo{journal}{Journal of management information systems} \bibinfo{volume}{19}, \bibinfo{pages}{9--30}.
%Type = Inproceedings
\bibitem[{Deng et~al.(2023)Deng, Xia, Peng, Yang and Zhang}]{deng2023large}
\bibinfo{author}{Deng, Y.}, \bibinfo{author}{Xia, C.S.}, \bibinfo{author}{Peng, H.}, \bibinfo{author}{Yang, C.}, \bibinfo{author}{Zhang, L.}, \bibinfo{year}{2023}.
\newblock \bibinfo{title}{Large language models are zero-shot fuzzers: Fuzzing deep-learning libraries via large language models}, in: \bibinfo{booktitle}{Proceedings of the 32nd ACM SIGSOFT international symposium on software testing and analysis}, pp. \bibinfo{pages}{423--435}.
%Type = Inproceedings
\bibitem[{Deng et~al.(2024)Deng, Xia, Yang, Zhang, Yang and Zhang}]{deng2024large}
\bibinfo{author}{Deng, Y.}, \bibinfo{author}{Xia, C.S.}, \bibinfo{author}{Yang, C.}, \bibinfo{author}{Zhang, S.D.}, \bibinfo{author}{Yang, S.}, \bibinfo{author}{Zhang, L.}, \bibinfo{year}{2024}.
\newblock \bibinfo{title}{Large language models are edge-case generators: Crafting unusual programs for fuzzing deep learning libraries}, in: \bibinfo{booktitle}{Proceedings of the 46th IEEE/ACM International Conference on Software Engineering}, pp. \bibinfo{pages}{1--13}.
%Type = Article
\bibitem[{Edstrom(1977)}]{edstrom1977user}
\bibinfo{author}{Edstrom, A.}, \bibinfo{year}{1977}.
\newblock \bibinfo{title}{User influence and the success of mis projects: a contingency approach}.
\newblock \bibinfo{journal}{Human Relations} \bibinfo{volume}{30}, \bibinfo{pages}{589--607}.
%Type = Inproceedings
\bibitem[{Feng and Chen(2024)}]{feng2024prompting}
\bibinfo{author}{Feng, S.}, \bibinfo{author}{Chen, C.}, \bibinfo{year}{2024}.
\newblock \bibinfo{title}{Prompting is all you need: Automated android bug replay with large language models}, in: \bibinfo{booktitle}{Proceedings of the 46th IEEE/ACM International Conference on Software Engineering}, pp. \bibinfo{pages}{1--13}.
%Type = Inproceedings
\bibitem[{Ferguson-Walter et~al.(2021)Ferguson-Walter, Major, Johnson and Muhleman}]{ferguson2021examining}
\bibinfo{author}{Ferguson-Walter, K.J.}, \bibinfo{author}{Major, M.M.}, \bibinfo{author}{Johnson, C.K.}, \bibinfo{author}{Muhleman, D.H.}, \bibinfo{year}{2021}.
\newblock \bibinfo{title}{Examining the efficacy of decoy-based and psychological cyber deception}, in: \bibinfo{booktitle}{30th USENIX security symposium (USENIX Security 21)}, pp. \bibinfo{pages}{1127--1144}.
%Type = Inproceedings
\bibitem[{van Gemert et~al.(2023)van Gemert, Hornb{\ae}k, Knibbe and Bergstr{\"o}m}]{van2023towards}
\bibinfo{author}{van Gemert, T.}, \bibinfo{author}{Hornb{\ae}k, K.}, \bibinfo{author}{Knibbe, J.}, \bibinfo{author}{Bergstr{\"o}m, J.}, \bibinfo{year}{2023}.
\newblock \bibinfo{title}{Towards a bedder future: A study of using virtual reality while lying down}, in: \bibinfo{booktitle}{Proceedings of the 2023 CHI Conference on Human Factors in Computing Systems}, pp. \bibinfo{pages}{1--18}.
%Type = Inproceedings
\bibitem[{Ghazi et~al.(2020)Ghazi, Kumar, Manurangsi and Pagh}]{ghazi2020private}
\bibinfo{author}{Ghazi, B.}, \bibinfo{author}{Kumar, R.}, \bibinfo{author}{Manurangsi, P.}, \bibinfo{author}{Pagh, R.}, \bibinfo{year}{2020}.
\newblock \bibinfo{title}{Private counting from anonymous messages: Near-optimal accuracy with vanishing communication overhead}, in: \bibinfo{booktitle}{International Conference on Machine Learning}, \bibinfo{organization}{PMLR}. pp. \bibinfo{pages}{3505--3514}.
%Type = Article
\bibitem[{Gu et~al.(2021)Gu, Huang, Hung and Chen}]{gu2021lessons}
\bibinfo{author}{Gu, H.}, \bibinfo{author}{Huang, J.}, \bibinfo{author}{Hung, L.}, \bibinfo{author}{Chen, X.}, \bibinfo{year}{2021}.
\newblock \bibinfo{title}{Lessons learned from designing an ai-enabled diagnosis tool for pathologists}.
\newblock \bibinfo{journal}{Proceedings of the ACM on Human-Computer Interaction} \bibinfo{volume}{5}, \bibinfo{pages}{1--25}.
%Type = Inproceedings
\bibitem[{Haimson et~al.(2023)Haimson, Nham, Thach and DeGuia}]{haimson2023transgender}
\bibinfo{author}{Haimson, O.L.}, \bibinfo{author}{Nham, K.}, \bibinfo{author}{Thach, H.}, \bibinfo{author}{DeGuia, A.}, \bibinfo{year}{2023}.
\newblock \bibinfo{title}{How transgender people and communities were involved in trans technology design processes}, in: \bibinfo{booktitle}{Proceedings of the 2023 CHI Conference on Human Factors in Computing Systems}, pp. \bibinfo{pages}{1--16}.
%Type = Inproceedings
\bibitem[{H{\"a}m{\"a}l{\"a}inen et~al.(2023)H{\"a}m{\"a}l{\"a}inen, Tavast and Kunnari}]{hamalainen2023evaluating}
\bibinfo{author}{H{\"a}m{\"a}l{\"a}inen, P.}, \bibinfo{author}{Tavast, M.}, \bibinfo{author}{Kunnari, A.}, \bibinfo{year}{2023}.
\newblock \bibinfo{title}{Evaluating large language models in generating synthetic hci research data: a case study}, in: \bibinfo{booktitle}{Proceedings of the 2023 CHI Conference on Human Factors in Computing Systems}, pp. \bibinfo{pages}{1--19}.
%Type = Article
\bibitem[{Hartono(2005)}]{hartono2005analisis}
\bibinfo{author}{Hartono, J.}, \bibinfo{year}{2005}.
\newblock \bibinfo{title}{Analisis dan desain sistem informasi: pendekatan terstruktur teori dan praktek aplikasi bisnis}.
\newblock \bibinfo{journal}{Yogyakarta: Andi} .
%Type = Article
\bibitem[{He and King(2008)}]{he2008role}
\bibinfo{author}{He, J.}, \bibinfo{author}{King, W.R.}, \bibinfo{year}{2008}.
\newblock \bibinfo{title}{The role of user participation in information systems development: implications from a meta-analysis}.
\newblock \bibinfo{journal}{Journal of management information systems} \bibinfo{volume}{25}, \bibinfo{pages}{301--331}.
%Type = Article
\bibitem[{He et~al.(2020)He, Ash, Anwar, Li, Yuan, Xu and Tian}]{he2020improving}
\bibinfo{author}{He, W.}, \bibinfo{author}{Ash, I.}, \bibinfo{author}{Anwar, M.}, \bibinfo{author}{Li, L.}, \bibinfo{author}{Yuan, X.}, \bibinfo{author}{Xu, L.}, \bibinfo{author}{Tian, X.}, \bibinfo{year}{2020}.
\newblock \bibinfo{title}{Improving employees’ intellectual capacity for cybersecurity through evidence-based malware training}.
\newblock \bibinfo{journal}{Journal of intellectual capital} \bibinfo{volume}{21}, \bibinfo{pages}{203--213}.
%Type = Article
\bibitem[{Hornb{\ae}k and Hertzum(2017)}]{hornbaek2017technology}
\bibinfo{author}{Hornb{\ae}k, K.}, \bibinfo{author}{Hertzum, M.}, \bibinfo{year}{2017}.
\newblock \bibinfo{title}{Technology acceptance and user experience: A review of the experiential component in hci}.
\newblock \bibinfo{journal}{ACM Transactions on Computer-Human Interaction (TOCHI)} \bibinfo{volume}{24}, \bibinfo{pages}{1--30}.
%Type = Inproceedings
\bibitem[{Huang et~al.(2023)Huang, Meng, Zhang, Liu, Wang, Li and Zhang}]{huang2023empirical}
\bibinfo{author}{Huang, K.}, \bibinfo{author}{Meng, X.}, \bibinfo{author}{Zhang, J.}, \bibinfo{author}{Liu, Y.}, \bibinfo{author}{Wang, W.}, \bibinfo{author}{Li, S.}, \bibinfo{author}{Zhang, Y.}, \bibinfo{year}{2023}.
\newblock \bibinfo{title}{An empirical study on fine-tuning large language models of code for automated program repair}, in: \bibinfo{booktitle}{2023 38th IEEE/ACM International Conference on Automated Software Engineering (ASE)}, \bibinfo{organization}{IEEE}. pp. \bibinfo{pages}{1162--1174}.
%Type = Article
\bibitem[{Huang and Pearlson(2019)}]{huang2019technology}
\bibinfo{author}{Huang, K.}, \bibinfo{author}{Pearlson, K.}, \bibinfo{year}{2019}.
\newblock \bibinfo{title}{For what technology can’t fix: Building a model of organizational cybersecurity culture} .
%Type = Article
\bibitem[{Huang et~al.(2022)Huang, Zhou and Chen}]{huang2022being}
\bibinfo{author}{Huang, K.}, \bibinfo{author}{Zhou, J.}, \bibinfo{author}{Chen, S.}, \bibinfo{year}{2022}.
\newblock \bibinfo{title}{Being a solo endeavor or team worker in crowdsourcing contests? it is a long-term decision you need to make}.
\newblock \bibinfo{journal}{Proceedings of the ACM on Human-Computer Interaction} \bibinfo{volume}{6}, \bibinfo{pages}{1--32}.
%Type = Article
\bibitem[{Iannone et~al.(2022)Iannone, Guadagni, Ferrucci, De~Lucia and Palomba}]{iannone2022secret}
\bibinfo{author}{Iannone, E.}, \bibinfo{author}{Guadagni, R.}, \bibinfo{author}{Ferrucci, F.}, \bibinfo{author}{De~Lucia, A.}, \bibinfo{author}{Palomba, F.}, \bibinfo{year}{2022}.
\newblock \bibinfo{title}{The secret life of software vulnerabilities: A large-scale empirical study}.
\newblock \bibinfo{journal}{IEEE Transactions on Software Engineering} \bibinfo{volume}{49}, \bibinfo{pages}{44--63}.
%Type = Article
\bibitem[{Iivari(2005)}]{iivari2005empirical}
\bibinfo{author}{Iivari, J.}, \bibinfo{year}{2005}.
\newblock \bibinfo{title}{An empirical test of the delone-mclean model of information system success}.
\newblock \bibinfo{journal}{ACM SIGMIS Database: the DATABASE for Advances in Information Systems} \bibinfo{volume}{36}, \bibinfo{pages}{8--27}.
%Type = Article
\bibitem[{Ives and Olson(1984)}]{ives1984user}
\bibinfo{author}{Ives, B.}, \bibinfo{author}{Olson, M.H.}, \bibinfo{year}{1984}.
\newblock \bibinfo{title}{User involvement and mis success: A review of research}.
\newblock \bibinfo{journal}{Management science} \bibinfo{volume}{30}, \bibinfo{pages}{586--603}.
%Type = Inproceedings
\bibitem[{Jacobs et~al.(2022)Jacobs, Beltiukov, Willinger, Ferreira, Gupta and Granville}]{jacobs2022ai}
\bibinfo{author}{Jacobs, A.S.}, \bibinfo{author}{Beltiukov, R.}, \bibinfo{author}{Willinger, W.}, \bibinfo{author}{Ferreira, R.A.}, \bibinfo{author}{Gupta, A.}, \bibinfo{author}{Granville, L.Z.}, \bibinfo{year}{2022}.
\newblock \bibinfo{title}{Ai/ml for network security: The emperor has no clothes}, in: \bibinfo{booktitle}{Proceedings of the 2022 ACM SIGSAC Conference on Computer and Communications Security}, pp. \bibinfo{pages}{1537--1551}.
%Type = Article
\bibitem[{Jacobs et~al.(2021)Jacobs, Pradier, McCoy~Jr, Perlis, Doshi-Velez and Gajos}]{jacobs2021machine}
\bibinfo{author}{Jacobs, M.}, \bibinfo{author}{Pradier, M.F.}, \bibinfo{author}{McCoy~Jr, T.H.}, \bibinfo{author}{Perlis, R.H.}, \bibinfo{author}{Doshi-Velez, F.}, \bibinfo{author}{Gajos, K.Z.}, \bibinfo{year}{2021}.
\newblock \bibinfo{title}{How machine-learning recommendations influence clinician treatment selections: the example of antidepressant selection}.
\newblock \bibinfo{journal}{Translational psychiatry} \bibinfo{volume}{11}, \bibinfo{pages}{108}.
%Type = Article
\bibitem[{Jeyaraj(2020)}]{jeyaraj2020delone}
\bibinfo{author}{Jeyaraj, A.}, \bibinfo{year}{2020}.
\newblock \bibinfo{title}{Delone \& mclean models of information system success: Critical meta-review and research directions}.
\newblock \bibinfo{journal}{International Journal of Information Management} \bibinfo{volume}{54}, \bibinfo{pages}{102139}.
%Type = Article
\bibitem[{Jia et~al.(2018)Jia, Qi, Shang, Jiang and Li}]{jia2018practical}
\bibinfo{author}{Jia, Y.}, \bibinfo{author}{Qi, Y.}, \bibinfo{author}{Shang, H.}, \bibinfo{author}{Jiang, R.}, \bibinfo{author}{Li, A.}, \bibinfo{year}{2018}.
\newblock \bibinfo{title}{A practical approach to constructing a knowledge graph for cybersecurity}.
\newblock \bibinfo{journal}{Engineering} \bibinfo{volume}{4}, \bibinfo{pages}{53--60}.
%Type = Inproceedings
\bibitem[{Jiang et~al.(2021)Jiang, Toh, Molina, Donsbach, Cai and Terry}]{jiang2021genline}
\bibinfo{author}{Jiang, E.}, \bibinfo{author}{Toh, E.}, \bibinfo{author}{Molina, A.}, \bibinfo{author}{Donsbach, A.}, \bibinfo{author}{Cai, C.J.}, \bibinfo{author}{Terry, M.}, \bibinfo{year}{2021}.
\newblock \bibinfo{title}{Genline and genform: Two tools for interacting with generative language models in a code editor}, in: \bibinfo{booktitle}{Adjunct Proceedings of the 34th Annual ACM Symposium on User Interface Software and Technology}, pp. \bibinfo{pages}{145--147}.
%Type = Inproceedings
\bibitem[{Jiang et~al.(2022)Jiang, Toh, Molina, Olson, Kayacik, Donsbach, Cai and Terry}]{jiang2022discovering}
\bibinfo{author}{Jiang, E.}, \bibinfo{author}{Toh, E.}, \bibinfo{author}{Molina, A.}, \bibinfo{author}{Olson, K.}, \bibinfo{author}{Kayacik, C.}, \bibinfo{author}{Donsbach, A.}, \bibinfo{author}{Cai, C.J.}, \bibinfo{author}{Terry, M.}, \bibinfo{year}{2022}.
\newblock \bibinfo{title}{Discovering the syntax and strategies of natural language programming with generative language models}, in: \bibinfo{booktitle}{Proceedings of the 2022 CHI Conference on Human Factors in Computing Systems}, pp. \bibinfo{pages}{1--19}.
%Type = Article
\bibitem[{Johnson and Whang(2002)}]{johnson2002business}
\bibinfo{author}{Johnson, M.E.}, \bibinfo{author}{Whang, S.}, \bibinfo{year}{2002}.
\newblock \bibinfo{title}{E-business and supply chain management: an overview and framework}.
\newblock \bibinfo{journal}{Production and Operations management} \bibinfo{volume}{11}, \bibinfo{pages}{413--423}.
%Type = Inproceedings
\bibitem[{Jones et~al.(2023)Jones, Neumayer and Shklovski}]{jones2023embodying}
\bibinfo{author}{Jones, M.}, \bibinfo{author}{Neumayer, C.}, \bibinfo{author}{Shklovski, I.}, \bibinfo{year}{2023}.
\newblock \bibinfo{title}{Embodying the algorithm: Exploring relationships with large language models through artistic performance}, in: \bibinfo{booktitle}{Proceedings of the 2023 CHI Conference on Human Factors in Computing Systems}, pp. \bibinfo{pages}{1--24}.
%Type = Inproceedings
\bibitem[{Kawakami et~al.(2022)Kawakami, Sivaraman, Cheng, Stapleton, Cheng, Qing, Perer, Wu, Zhu and Holstein}]{kawakami2022improving}
\bibinfo{author}{Kawakami, A.}, \bibinfo{author}{Sivaraman, V.}, \bibinfo{author}{Cheng, H.F.}, \bibinfo{author}{Stapleton, L.}, \bibinfo{author}{Cheng, Y.}, \bibinfo{author}{Qing, D.}, \bibinfo{author}{Perer, A.}, \bibinfo{author}{Wu, Z.S.}, \bibinfo{author}{Zhu, H.}, \bibinfo{author}{Holstein, K.}, \bibinfo{year}{2022}.
\newblock \bibinfo{title}{Improving human-ai partnerships in child welfare: understanding worker practices, challenges, and desires for algorithmic decision support}, in: \bibinfo{booktitle}{Proceedings of the 2022 CHI Conference on Human Factors in Computing Systems}, pp. \bibinfo{pages}{1--18}.
%Type = Inproceedings
\bibitem[{Kazemitabaar et~al.(2023)Kazemitabaar, Chow, Ma, Ericson, Weintrop and Grossman}]{kazemitabaar2023studying}
\bibinfo{author}{Kazemitabaar, M.}, \bibinfo{author}{Chow, J.}, \bibinfo{author}{Ma, C.K.T.}, \bibinfo{author}{Ericson, B.J.}, \bibinfo{author}{Weintrop, D.}, \bibinfo{author}{Grossman, T.}, \bibinfo{year}{2023}.
\newblock \bibinfo{title}{Studying the effect of ai code generators on supporting novice learners in introductory programming}, in: \bibinfo{booktitle}{Proceedings of the 2023 CHI Conference on Human Factors in Computing Systems}, pp. \bibinfo{pages}{1--23}.
%Type = Article
\bibitem[{Khairat et~al.(2018)Khairat, Marc, Crosby, Al~Sanousi et~al.}]{khairat2018reasons}
\bibinfo{author}{Khairat, S.}, \bibinfo{author}{Marc, D.}, \bibinfo{author}{Crosby, W.}, \bibinfo{author}{Al~Sanousi, A.}, et~al., \bibinfo{year}{2018}.
\newblock \bibinfo{title}{Reasons for physicians not adopting clinical decision support systems: critical analysis}.
\newblock \bibinfo{journal}{JMIR medical informatics} \bibinfo{volume}{6}, \bibinfo{pages}{e8912}.
%Type = Inproceedings
\bibitem[{Kizilcec(2016)}]{kizilcec2016much}
\bibinfo{author}{Kizilcec, R.F.}, \bibinfo{year}{2016}.
\newblock \bibinfo{title}{How much information? effects of transparency on trust in an algorithmic interface}, in: \bibinfo{booktitle}{Proceedings of the 2016 CHI conference on human factors in computing systems}, pp. \bibinfo{pages}{2390--2395}.
%Type = Inproceedings
\bibitem[{Kokulu et~al.(2019)Kokulu, Soneji, Bao, Shoshitaishvili, Zhao, Doup{\'e} and Ahn}]{kokulu2019matched}
\bibinfo{author}{Kokulu, F.B.}, \bibinfo{author}{Soneji, A.}, \bibinfo{author}{Bao, T.}, \bibinfo{author}{Shoshitaishvili, Y.}, \bibinfo{author}{Zhao, Z.}, \bibinfo{author}{Doup{\'e}, A.}, \bibinfo{author}{Ahn, G.J.}, \bibinfo{year}{2019}.
\newblock \bibinfo{title}{Matched and mismatched socs: A qualitative study on security operations center issues}, in: \bibinfo{booktitle}{Proceedings of the 2019 ACM SIGSAC conference on computer and communications security}, pp. \bibinfo{pages}{1955--1970}.
%Type = Article
\bibitem[{Kreps et~al.(2022)Kreps, McCain and Brundage}]{kreps2022all}
\bibinfo{author}{Kreps, S.}, \bibinfo{author}{McCain, R.M.}, \bibinfo{author}{Brundage, M.}, \bibinfo{year}{2022}.
\newblock \bibinfo{title}{All the news that’s fit to fabricate: Ai-generated text as a tool of media misinformation}.
\newblock \bibinfo{journal}{Journal of experimental political science} \bibinfo{volume}{9}, \bibinfo{pages}{104--117}.
%Type = Inproceedings
\bibitem[{Ku{\v{c}}era et~al.(2021)Ku{\v{c}}era, Scott, Lindley and Olivier}]{kuvcera2021bedtime}
\bibinfo{author}{Ku{\v{c}}era, J.}, \bibinfo{author}{Scott, J.}, \bibinfo{author}{Lindley, S.}, \bibinfo{author}{Olivier, P.}, \bibinfo{year}{2021}.
\newblock \bibinfo{title}{Bedtime window: A field study connecting bedrooms of long-distance couples using a slow photo-stream and shared real-time inking}, in: \bibinfo{booktitle}{Proceedings of the 2021 CHI Conference on Human Factors in Computing Systems}, pp. \bibinfo{pages}{1--12}.
%Type = Article
\bibitem[{Kujala(2003)}]{kujala2003user}
\bibinfo{author}{Kujala, S.}, \bibinfo{year}{2003}.
\newblock \bibinfo{title}{User involvement: a review of the benefits and challenges}.
\newblock \bibinfo{journal}{Behaviour \& information technology} \bibinfo{volume}{22}, \bibinfo{pages}{1--16}.
%Type = Article
\bibitem[{Kumar and Ros{\'e}(2014)}]{kumar2014triggering}
\bibinfo{author}{Kumar, R.}, \bibinfo{author}{Ros{\'e}, C.P.}, \bibinfo{year}{2014}.
\newblock \bibinfo{title}{Triggering effective social support for online groups}.
\newblock \bibinfo{journal}{ACM Transactions on Interactive Intelligent Systems (TiiS)} \bibinfo{volume}{3}, \bibinfo{pages}{1--32}.
%Type = Article
\bibitem[{Kweon et~al.(2021)Kweon, Lee, Chai and Yoo}]{kweon2021utility}
\bibinfo{author}{Kweon, E.}, \bibinfo{author}{Lee, H.}, \bibinfo{author}{Chai, S.}, \bibinfo{author}{Yoo, K.}, \bibinfo{year}{2021}.
\newblock \bibinfo{title}{The utility of information security training and education on cybersecurity incidents: An empirical evidence}.
\newblock \bibinfo{journal}{Information Systems Frontiers} \bibinfo{volume}{23}, \bibinfo{pages}{361--373}.
%Type = Inproceedings
\bibitem[{Lai et~al.(2022)Lai, Carton, Bhatnagar, Liao, Zhang and Tan}]{lai2022human}
\bibinfo{author}{Lai, V.}, \bibinfo{author}{Carton, S.}, \bibinfo{author}{Bhatnagar, R.}, \bibinfo{author}{Liao, Q.V.}, \bibinfo{author}{Zhang, Y.}, \bibinfo{author}{Tan, C.}, \bibinfo{year}{2022}.
\newblock \bibinfo{title}{Human-ai collaboration via conditional delegation: A case study of content moderation}, in: \bibinfo{booktitle}{Proceedings of the 2022 CHI Conference on Human Factors in Computing Systems}, pp. \bibinfo{pages}{1--18}.
%Type = Article
\bibitem[{Lewandowska et~al.(2018)Lewandowska, Wachowicz, Marek, Oginska and Fafrowicz}]{lewandowska2018would}
\bibinfo{author}{Lewandowska, K.}, \bibinfo{author}{Wachowicz, B.}, \bibinfo{author}{Marek, T.}, \bibinfo{author}{Oginska, H.}, \bibinfo{author}{Fafrowicz, M.}, \bibinfo{year}{2018}.
\newblock \bibinfo{title}{Would you say “yes” in the evening? time-of-day effect on response bias in four types of working memory recognition tasks}.
\newblock \bibinfo{journal}{Chronobiology international} \bibinfo{volume}{35}, \bibinfo{pages}{80--89}.
%Type = Article
\bibitem[{Li et~al.(2019)Li, He, Xu, Ash, Anwar and Yuan}]{li2019investigating}
\bibinfo{author}{Li, L.}, \bibinfo{author}{He, W.}, \bibinfo{author}{Xu, L.}, \bibinfo{author}{Ash, I.}, \bibinfo{author}{Anwar, M.}, \bibinfo{author}{Yuan, X.}, \bibinfo{year}{2019}.
\newblock \bibinfo{title}{Investigating the impact of cybersecurity policy awareness on employees’ cybersecurity behavior}.
\newblock \bibinfo{journal}{International Journal of Information Management} \bibinfo{volume}{45}, \bibinfo{pages}{13--24}.
%Type = Inproceedings
\bibitem[{Liang et~al.(2019)Liang, Proft, Andersen and Knepper}]{liang2019implicit}
\bibinfo{author}{Liang, C.}, \bibinfo{author}{Proft, J.}, \bibinfo{author}{Andersen, E.}, \bibinfo{author}{Knepper, R.A.}, \bibinfo{year}{2019}.
\newblock \bibinfo{title}{Implicit communication of actionable information in human-ai teams}, in: \bibinfo{booktitle}{Proceedings of the 2019 CHI Conference on Human Factors in Computing Systems}, pp. \bibinfo{pages}{1--13}.
%Type = Inproceedings
\bibitem[{Liu et~al.(2023)Liu, Sarkar, Negreanu, Zorn, Williams, Toronto and Gordon}]{liu2023wants}
\bibinfo{author}{Liu, M.X.}, \bibinfo{author}{Sarkar, A.}, \bibinfo{author}{Negreanu, C.}, \bibinfo{author}{Zorn, B.}, \bibinfo{author}{Williams, J.}, \bibinfo{author}{Toronto, N.}, \bibinfo{author}{Gordon, A.D.}, \bibinfo{year}{2023}.
\newblock \bibinfo{title}{“what it wants me to say”: Bridging the abstraction gap between end-user programmers and code-generating large language models}, in: \bibinfo{booktitle}{Proceedings of the 2023 CHI Conference on Human Factors in Computing Systems}, pp. \bibinfo{pages}{1--31}.
%Type = Inproceedings
\bibitem[{Lurie and Mulligan(2020)}]{lurie2020crowdworkers}
\bibinfo{author}{Lurie, E.}, \bibinfo{author}{Mulligan, D.K.}, \bibinfo{year}{2020}.
\newblock \bibinfo{title}{Crowdworkers are not judges: Rethinking crowdsourced vignette studies as a risk assessment evaluation technique}, in: \bibinfo{booktitle}{Proceedings of the Workshop on Fair and Responsible AI at CHI}.
%Type = Article
\bibitem[{Mahdavifar and Ghorbani(2020)}]{mahdavifar2020dennes}
\bibinfo{author}{Mahdavifar, S.}, \bibinfo{author}{Ghorbani, A.A.}, \bibinfo{year}{2020}.
\newblock \bibinfo{title}{Dennes: deep embedded neural network expert system for detecting cyber attacks}.
\newblock \bibinfo{journal}{Neural Computing and Applications} \bibinfo{volume}{32}, \bibinfo{pages}{14753--14780}.
%Type = Article
\bibitem[{Marek et~al.(2015)Marek, Brock and Savla}]{marek2015evaluating}
\bibinfo{author}{Marek, L.I.}, \bibinfo{author}{Brock, D.J.P.}, \bibinfo{author}{Savla, J.}, \bibinfo{year}{2015}.
\newblock \bibinfo{title}{Evaluating collaboration for effectiveness: Conceptualization and measurement}.
\newblock \bibinfo{journal}{American Journal of Evaluation} \bibinfo{volume}{36}, \bibinfo{pages}{67--85}.
%Type = Article
\bibitem[{McKeen et~al.(1994)McKeen, Guimaraes and Wetherbe}]{mckeen1994relationship}
\bibinfo{author}{McKeen, J.D.}, \bibinfo{author}{Guimaraes, T.}, \bibinfo{author}{Wetherbe, J.C.}, \bibinfo{year}{1994}.
\newblock \bibinfo{title}{The relationship between user participation and user satisfaction: an investigation of four contingency factors}.
\newblock \bibinfo{journal}{MIS quarterly} , \bibinfo{pages}{427--451}.
%Type = Inproceedings
\bibitem[{McNutt et~al.(2023)McNutt, Wang, Deline and Drucker}]{mcnutt2023design}
\bibinfo{author}{McNutt, A.M.}, \bibinfo{author}{Wang, C.}, \bibinfo{author}{Deline, R.A.}, \bibinfo{author}{Drucker, S.M.}, \bibinfo{year}{2023}.
\newblock \bibinfo{title}{On the design of ai-powered code assistants for notebooks}, in: \bibinfo{booktitle}{Proceedings of the 2023 CHI Conference on Human Factors in Computing Systems}, pp. \bibinfo{pages}{1--16}.
%Type = Inproceedings
\bibitem[{Mendsaikhan et~al.(2019)Mendsaikhan, Hasegawa, Yamaguchi and Shimada}]{mendsaikhan2019identification}
\bibinfo{author}{Mendsaikhan, O.}, \bibinfo{author}{Hasegawa, H.}, \bibinfo{author}{Yamaguchi, Y.}, \bibinfo{author}{Shimada, H.}, \bibinfo{year}{2019}.
\newblock \bibinfo{title}{Identification of cybersecurity specific content using the doc2vec language model}, in: \bibinfo{booktitle}{2019 IEEE 43rd annual computer software and applications conference (COMPSAC)}, \bibinfo{organization}{IEEE}. pp. \bibinfo{pages}{396--401}.
%Type = Inproceedings
\bibitem[{Meng et~al.(2024)Meng, Mirchev, B{\"o}hme and Roychoudhury}]{meng2024large}
\bibinfo{author}{Meng, R.}, \bibinfo{author}{Mirchev, M.}, \bibinfo{author}{B{\"o}hme, M.}, \bibinfo{author}{Roychoudhury, A.}, \bibinfo{year}{2024}.
\newblock \bibinfo{title}{Large language model guided protocol fuzzing}, in: \bibinfo{booktitle}{Proceedings of the 31st Annual Network and Distributed System Security Symposium (NDSS)}.
%Type = Inproceedings
\bibitem[{Mirowski et~al.(2023)Mirowski, Mathewson, Pittman and Evans}]{mirowski2023co}
\bibinfo{author}{Mirowski, P.}, \bibinfo{author}{Mathewson, K.W.}, \bibinfo{author}{Pittman, J.}, \bibinfo{author}{Evans, R.}, \bibinfo{year}{2023}.
\newblock \bibinfo{title}{Co-writing screenplays and theatre scripts with language models: Evaluation by industry professionals}, in: \bibinfo{booktitle}{Proceedings of the 2023 CHI Conference on Human Factors in Computing Systems}, pp. \bibinfo{pages}{1--34}.
%Type = Article
\bibitem[{Mitchell et~al.(2021)Mitchell, Maimone, Cassells, Tobin, Davidson, Smaldone and Mamykina}]{mitchell2021automated}
\bibinfo{author}{Mitchell, E.G.}, \bibinfo{author}{Maimone, R.}, \bibinfo{author}{Cassells, A.}, \bibinfo{author}{Tobin, J.N.}, \bibinfo{author}{Davidson, P.}, \bibinfo{author}{Smaldone, A.M.}, \bibinfo{author}{Mamykina, L.}, \bibinfo{year}{2021}.
\newblock \bibinfo{title}{Automated vs. human health coaching: exploring participant and practitioner experiences}.
\newblock \bibinfo{journal}{Proceedings of the ACM on human-computer interaction} \bibinfo{volume}{5}, \bibinfo{pages}{1--37}.
%Type = Article
\bibitem[{Mohammed~Noori and Ozdamli(2022)}]{mohammed2022evaluating}
\bibinfo{author}{Mohammed~Noori, N.}, \bibinfo{author}{Ozdamli, F.}, \bibinfo{year}{2022}.
\newblock \bibinfo{title}{Evaluating e-learning system success in higher education during the covid-19}.
\newblock \bibinfo{journal}{Cypriot Journal of Educational Sciences} \bibinfo{volume}{17}, \bibinfo{pages}{4884--4913}.
%Type = Inproceedings
\bibitem[{Muresan and Pohl(2019)}]{muresan2019chats}
\bibinfo{author}{Muresan, A.}, \bibinfo{author}{Pohl, H.}, \bibinfo{year}{2019}.
\newblock \bibinfo{title}{Chats with bots: Balancing imitation and engagement}, in: \bibinfo{booktitle}{Extended abstracts of the 2019 CHI conference on human factors in computing systems}, pp. \bibinfo{pages}{1--6}.
%Type = Inproceedings
\bibitem[{Nicholson et~al.(2019)Nicholson, Coventry and Briggs}]{nicholson2019if}
\bibinfo{author}{Nicholson, J.}, \bibinfo{author}{Coventry, L.}, \bibinfo{author}{Briggs, P.}, \bibinfo{year}{2019}.
\newblock \bibinfo{title}{" if it's important it will be a headline" cybersecurity information seeking in older adults}, in: \bibinfo{booktitle}{Proceedings of the 2019 CHI Conference on Human Factors in Computing Systems}, pp. \bibinfo{pages}{1--11}.
%Type = Inproceedings
\bibitem[{OpenAI(2022)}]{ChatGPT2022}
\bibinfo{author}{OpenAI}, \bibinfo{year}{2022}.
\newblock \bibinfo{title}{Chatgpt: Optimizing language models for dialogue.}
\newblock \URLprefix \url{https: //openai.com/blog/chatgpt/}.
%Type = Article
\bibitem[{OpenAI(2023)}]{openai2023gpt}
\bibinfo{author}{OpenAI}, \bibinfo{year}{2023}.
\newblock \bibinfo{title}{Gpt-4 technical report. arxiv 2303.08774} .
%Type = Inproceedings
\bibitem[{Panigutti et~al.(2022)Panigutti, Beretta, Giannotti and Pedreschi}]{panigutti2022understanding}
\bibinfo{author}{Panigutti, C.}, \bibinfo{author}{Beretta, A.}, \bibinfo{author}{Giannotti, F.}, \bibinfo{author}{Pedreschi, D.}, \bibinfo{year}{2022}.
\newblock \bibinfo{title}{Understanding the impact of explanations on advice-taking: a user study for ai-based clinical decision support systems}, in: \bibinfo{booktitle}{Proceedings of the 2022 CHI Conference on Human Factors in Computing Systems}, pp. \bibinfo{pages}{1--9}.
%Type = Inproceedings
\bibitem[{Pearce et~al.(2023)Pearce, Tan, Ahmad, Karri and Dolan-Gavitt}]{pearce2023examining}
\bibinfo{author}{Pearce, H.}, \bibinfo{author}{Tan, B.}, \bibinfo{author}{Ahmad, B.}, \bibinfo{author}{Karri, R.}, \bibinfo{author}{Dolan-Gavitt, B.}, \bibinfo{year}{2023}.
\newblock \bibinfo{title}{Examining zero-shot vulnerability repair with large language models}, in: \bibinfo{booktitle}{2023 IEEE Symposium on Security and Privacy (SP)}, \bibinfo{organization}{IEEE}. pp. \bibinfo{pages}{2339--2356}.
%Type = Article
\bibitem[{Pearlson and Huang(2022)}]{pearlson2022design}
\bibinfo{author}{Pearlson, K.}, \bibinfo{author}{Huang, K.}, \bibinfo{year}{2022}.
\newblock \bibinfo{title}{Design for cybersecurity from the start}.
\newblock \bibinfo{journal}{MIT Sloan Management Review} \bibinfo{volume}{63}, \bibinfo{pages}{73--77}.
%Type = Inproceedings
\bibitem[{Petridis et~al.(2023)Petridis, Diakopoulos, Crowston, Hansen, Henderson, Jastrzebski, Nickerson and Chilton}]{petridis2023anglekindling}
\bibinfo{author}{Petridis, S.}, \bibinfo{author}{Diakopoulos, N.}, \bibinfo{author}{Crowston, K.}, \bibinfo{author}{Hansen, M.}, \bibinfo{author}{Henderson, K.}, \bibinfo{author}{Jastrzebski, S.}, \bibinfo{author}{Nickerson, J.V.}, \bibinfo{author}{Chilton, L.B.}, \bibinfo{year}{2023}.
\newblock \bibinfo{title}{Anglekindling: Supporting journalistic angle ideation with large language models}, in: \bibinfo{booktitle}{Proceedings of the 2023 CHI Conference on Human Factors in Computing Systems}, pp. \bibinfo{pages}{1--16}.
%Type = Article
\bibitem[{Posey and Folger(2020)}]{posey2020exploratory}
\bibinfo{author}{Posey, C.}, \bibinfo{author}{Folger, R.}, \bibinfo{year}{2020}.
\newblock \bibinfo{title}{An exploratory examination of organizational insiders’ descriptive and normative perceptions of cyber-relevant rights and responsibilities}.
\newblock \bibinfo{journal}{Computers \& security} \bibinfo{volume}{99}, \bibinfo{pages}{102038}.
%Type = Article
\bibitem[{Riebe et~al.(2021)Riebe, Kaufhold and Reuter}]{riebe2021impact}
\bibinfo{author}{Riebe, T.}, \bibinfo{author}{Kaufhold, M.A.}, \bibinfo{author}{Reuter, C.}, \bibinfo{year}{2021}.
\newblock \bibinfo{title}{The impact of organizational structure and technology use on collaborative practices in computer emergency response teams: An empirical study}.
\newblock \bibinfo{journal}{Proceedings of the ACM on human-computer interaction} \bibinfo{volume}{5}, \bibinfo{pages}{1--30}.
%Type = Article
\bibitem[{Sejnowski(2023)}]{sejnowski2023large}
\bibinfo{author}{Sejnowski, T.J.}, \bibinfo{year}{2023}.
\newblock \bibinfo{title}{Large language models and the reverse turing test}.
\newblock \bibinfo{journal}{Neural computation} \bibinfo{volume}{35}, \bibinfo{pages}{309--342}.
%Type = Article
\bibitem[{Sewak et~al.(2023)Sewak, Sahay and Rathore}]{sewak2023deep}
\bibinfo{author}{Sewak, M.}, \bibinfo{author}{Sahay, S.K.}, \bibinfo{author}{Rathore, H.}, \bibinfo{year}{2023}.
\newblock \bibinfo{title}{Deep reinforcement learning in the advanced cybersecurity threat detection and protection}.
\newblock \bibinfo{journal}{Information Systems Frontiers} \bibinfo{volume}{25}, \bibinfo{pages}{589--611}.
%Type = Inproceedings
\bibitem[{Sivaraman et~al.(2023)Sivaraman, Bukowski, Levin, Kahn and Perer}]{sivaraman2023ignore}
\bibinfo{author}{Sivaraman, V.}, \bibinfo{author}{Bukowski, L.A.}, \bibinfo{author}{Levin, J.}, \bibinfo{author}{Kahn, J.M.}, \bibinfo{author}{Perer, A.}, \bibinfo{year}{2023}.
\newblock \bibinfo{title}{Ignore, trust, or negotiate: understanding clinician acceptance of ai-based treatment recommendations in health care}, in: \bibinfo{booktitle}{Proceedings of the 2023 CHI Conference on Human Factors in Computing Systems}, pp. \bibinfo{pages}{1--18}.
%Type = Article
\bibitem[{Stempfle and Badke-Schaub(2002)}]{stempfle2002thinking}
\bibinfo{author}{Stempfle, J.}, \bibinfo{author}{Badke-Schaub, P.}, \bibinfo{year}{2002}.
\newblock \bibinfo{title}{Thinking in design teams-an analysis of team communication}.
\newblock \bibinfo{journal}{Design studies} \bibinfo{volume}{23}, \bibinfo{pages}{473--496}.
%Type = Article
\bibitem[{Taddeo et~al.(2019)Taddeo, McCutcheon and Floridi}]{taddeo2019trusting}
\bibinfo{author}{Taddeo, M.}, \bibinfo{author}{McCutcheon, T.}, \bibinfo{author}{Floridi, L.}, \bibinfo{year}{2019}.
\newblock \bibinfo{title}{Trusting artificial intelligence in cybersecurity is a double-edged sword}.
\newblock \bibinfo{journal}{Nature Machine Intelligence} \bibinfo{volume}{1}, \bibinfo{pages}{557--560}.
%Type = Inproceedings
\bibitem[{Tan et~al.(2021)Tan, Yang, Al-Shedivat, Xing and Hu}]{tan2021progressive}
\bibinfo{author}{Tan, B.}, \bibinfo{author}{Yang, Z.}, \bibinfo{author}{Al-Shedivat, M.}, \bibinfo{author}{Xing, E.}, \bibinfo{author}{Hu, Z.}, \bibinfo{year}{2021}.
\newblock \bibinfo{title}{Progressive generation of long text with pretrained language models}, in: \bibinfo{booktitle}{Proceedings of the 2021 Conference of the North American Chapter of the Association for Computational Linguistics: Human Language Technologies}, pp. \bibinfo{pages}{4313--4324}.
%Type = Article
\bibitem[{Tarafdar et~al.(2010)Tarafdar, Tu and Ragu-Nathan}]{tarafdar2010impact}
\bibinfo{author}{Tarafdar, M.}, \bibinfo{author}{Tu, Q.}, \bibinfo{author}{Ragu-Nathan, T.}, \bibinfo{year}{2010}.
\newblock \bibinfo{title}{Impact of technostress on end-user satisfaction and performance}.
\newblock \bibinfo{journal}{Journal of management information systems} \bibinfo{volume}{27}, \bibinfo{pages}{303--334}.
%Type = Article
\bibitem[{Toet et~al.(2016)Toet, Brouwer, van~den Bosch and Korteling}]{toet2016effects}
\bibinfo{author}{Toet, A.}, \bibinfo{author}{Brouwer, A.M.}, \bibinfo{author}{van~den Bosch, K.}, \bibinfo{author}{Korteling, J.}, \bibinfo{year}{2016}.
\newblock \bibinfo{title}{Effects of personal characteristics on susceptibility to decision bias: a literature study}.
\newblock \bibinfo{journal}{International Journal of Humanities and Social Sciences} \bibinfo{volume}{8}, \bibinfo{pages}{1--17}.
%Type = Inproceedings
\bibitem[{Tsai et~al.(2021)Tsai, You, Gui, Kou and Carroll}]{tsai2021exploring}
\bibinfo{author}{Tsai, C.H.}, \bibinfo{author}{You, Y.}, \bibinfo{author}{Gui, X.}, \bibinfo{author}{Kou, Y.}, \bibinfo{author}{Carroll, J.M.}, \bibinfo{year}{2021}.
\newblock \bibinfo{title}{Exploring and promoting diagnostic transparency and explainability in online symptom checkers}, in: \bibinfo{booktitle}{Proceedings of the 2021 CHI Conference on Human Factors in Computing Systems}, pp. \bibinfo{pages}{1--17}.
%Type = Article
\bibitem[{Urbach et~al.(2010)Urbach, Smolnik and Riempp}]{urbach2010improving}
\bibinfo{author}{Urbach, N.}, \bibinfo{author}{Smolnik, S.}, \bibinfo{author}{Riempp, G.}, \bibinfo{year}{2010}.
\newblock \bibinfo{title}{Improving the success of employee portals: A causal and performance-based analysis} .
%Type = Article
\bibitem[{Varghese and Chapiro(2023)}]{varghese2023chatgpt}
\bibinfo{author}{Varghese, J.}, \bibinfo{author}{Chapiro, J.}, \bibinfo{year}{2023}.
\newblock \bibinfo{title}{Chatgpt: The transformative influence of generative ai on science and healthcare}.
\newblock \bibinfo{journal}{Journal of Hepatology} .
%Type = Inproceedings
\bibitem[{Verma et~al.(2023)Verma, Mlynar, Schaer, Reichenbach, Jreige, Prior, Ev{\'e}quoz and Depeursinge}]{verma2023rethinking}
\bibinfo{author}{Verma, H.}, \bibinfo{author}{Mlynar, J.}, \bibinfo{author}{Schaer, R.}, \bibinfo{author}{Reichenbach, J.}, \bibinfo{author}{Jreige, M.}, \bibinfo{author}{Prior, J.}, \bibinfo{author}{Ev{\'e}quoz, F.}, \bibinfo{author}{Depeursinge, A.}, \bibinfo{year}{2023}.
\newblock \bibinfo{title}{Rethinking the role of ai with physicians in oncology: revealing perspectives from clinical and research workflows}, in: \bibinfo{booktitle}{Proceedings of the 2023 CHI Conference on Human Factors in Computing Systems}, pp. \bibinfo{pages}{1--19}.
%Type = Article
\bibitem[{Victor~Chen et~al.(2013)Victor~Chen, Chen and Paolo S.~Capistrano}]{victor2013process}
\bibinfo{author}{Victor~Chen, J.}, \bibinfo{author}{Chen, Y.}, \bibinfo{author}{Paolo S.~Capistrano, E.}, \bibinfo{year}{2013}.
\newblock \bibinfo{title}{Process quality and collaboration quality on b2b e-commerce}.
\newblock \bibinfo{journal}{Industrial Management \& Data Systems} \bibinfo{volume}{113}, \bibinfo{pages}{908--926}.
%Type = Article
\bibitem[{Vitturi et~al.(2013)Vitturi, Tramarin and Seno}]{vitturi2013industrial}
\bibinfo{author}{Vitturi, S.}, \bibinfo{author}{Tramarin, F.}, \bibinfo{author}{Seno, L.}, \bibinfo{year}{2013}.
\newblock \bibinfo{title}{Industrial wireless networks: The significance of timeliness in communication systems}.
\newblock \bibinfo{journal}{IEEE Industrial Electronics Magazine} \bibinfo{volume}{7}, \bibinfo{pages}{40--51}.
%Type = Inproceedings
\bibitem[{Wang et~al.(2023)Wang, Petridis, Kwon, Ma and Chilton}]{wang2023popblends}
\bibinfo{author}{Wang, S.}, \bibinfo{author}{Petridis, S.}, \bibinfo{author}{Kwon, T.}, \bibinfo{author}{Ma, X.}, \bibinfo{author}{Chilton, L.B.}, \bibinfo{year}{2023}.
\newblock \bibinfo{title}{Popblends: Strategies for conceptual blending with large language models}, in: \bibinfo{booktitle}{Proceedings of the 2023 CHI Conference on Human Factors in Computing Systems}, pp. \bibinfo{pages}{1--19}.
%Type = Article
\bibitem[{Wei et~al.(2022)Wei, Wang, Schuurmans, Bosma, Xia, Chi, Le, Zhou et~al.}]{wei2022chain}
\bibinfo{author}{Wei, J.}, \bibinfo{author}{Wang, X.}, \bibinfo{author}{Schuurmans, D.}, \bibinfo{author}{Bosma, M.}, \bibinfo{author}{Xia, F.}, \bibinfo{author}{Chi, E.}, \bibinfo{author}{Le, Q.V.}, \bibinfo{author}{Zhou, D.}, et~al., \bibinfo{year}{2022}.
\newblock \bibinfo{title}{Chain-of-thought prompting elicits reasoning in large language models}.
\newblock \bibinfo{journal}{Advances in Neural Information Processing Systems} \bibinfo{volume}{35}, \bibinfo{pages}{24824--24837}.
%Type = Article
\bibitem[{White et~al.(2018)White, Clough and Casey}]{white2018help}
\bibinfo{author}{White, M.M.}, \bibinfo{author}{Clough, B.A.}, \bibinfo{author}{Casey, L.M.}, \bibinfo{year}{2018}.
\newblock \bibinfo{title}{What do help-seeking measures assess? building a conceptualization framework for help-seeking intentions through a systematic review of measure content}.
\newblock \bibinfo{journal}{Clinical psychology review} \bibinfo{volume}{59}, \bibinfo{pages}{61--77}.
%Type = Article
\bibitem[{Whyte et~al.(1997)Whyte, Bytheway and Edwards}]{whyte1997understanding}
\bibinfo{author}{Whyte, G.}, \bibinfo{author}{Bytheway, A.}, \bibinfo{author}{Edwards, C.}, \bibinfo{year}{1997}.
\newblock \bibinfo{title}{Understanding user perceptions of information systems success}.
\newblock \bibinfo{journal}{The Journal of Strategic Information Systems} \bibinfo{volume}{6}, \bibinfo{pages}{35--68}.
%Type = Article
\bibitem[{Wiczorek and Meyer(2019)}]{wiczorek2019effects}
\bibinfo{author}{Wiczorek, R.}, \bibinfo{author}{Meyer, J.}, \bibinfo{year}{2019}.
\newblock \bibinfo{title}{Effects of trust, self-confidence, and feedback on the use of decision automation}.
\newblock \bibinfo{journal}{Frontiers in psychology} \bibinfo{volume}{10}, \bibinfo{pages}{519}.
%Type = Article
\bibitem[{Wixom and Todd(2005)}]{wixom2005theoretical}
\bibinfo{author}{Wixom, B.H.}, \bibinfo{author}{Todd, P.A.}, \bibinfo{year}{2005}.
\newblock \bibinfo{title}{A theoretical integration of user satisfaction and technology acceptance}.
\newblock \bibinfo{journal}{Information systems research} \bibinfo{volume}{16}, \bibinfo{pages}{85--102}.
%Type = Inproceedings
\bibitem[{Wu et~al.(2022)Wu, Terry and Cai}]{wu2022ai}
\bibinfo{author}{Wu, T.}, \bibinfo{author}{Terry, M.}, \bibinfo{author}{Cai, C.J.}, \bibinfo{year}{2022}.
\newblock \bibinfo{title}{Ai chains: Transparent and controllable human-ai interaction by chaining large language model prompts}, in: \bibinfo{booktitle}{Proceedings of the 2022 CHI conference on human factors in computing systems}, pp. \bibinfo{pages}{1--22}.
%Type = Inproceedings
\bibitem[{Xia et~al.(2024)Xia, Paltenghi, Le~Tian, Pradel and Zhang}]{xia2024fuzz4all}
\bibinfo{author}{Xia, C.S.}, \bibinfo{author}{Paltenghi, M.}, \bibinfo{author}{Le~Tian, J.}, \bibinfo{author}{Pradel, M.}, \bibinfo{author}{Zhang, L.}, \bibinfo{year}{2024}.
\newblock \bibinfo{title}{Fuzz4all: Universal fuzzing with large language models}, in: \bibinfo{booktitle}{Proceedings of the IEEE/ACM 46th International Conference on Software Engineering}, pp. \bibinfo{pages}{1--13}.
%Type = Inproceedings
\bibitem[{Xia et~al.(2023)Xia, Wei and Zhang}]{xia2023automated}
\bibinfo{author}{Xia, C.S.}, \bibinfo{author}{Wei, Y.}, \bibinfo{author}{Zhang, L.}, \bibinfo{year}{2023}.
\newblock \bibinfo{title}{Automated program repair in the era of large pre-trained language models}, in: \bibinfo{booktitle}{2023 IEEE/ACM 45th International Conference on Software Engineering (ICSE)}, \bibinfo{organization}{IEEE}. pp. \bibinfo{pages}{1482--1494}.
%Type = Inproceedings
\bibitem[{Yang et~al.(2024)Yang, Le~Goues, Martins and Hellendoorn}]{yang2024large}
\bibinfo{author}{Yang, A.Z.}, \bibinfo{author}{Le~Goues, C.}, \bibinfo{author}{Martins, R.}, \bibinfo{author}{Hellendoorn, V.}, \bibinfo{year}{2024}.
\newblock \bibinfo{title}{Large language models for test-free fault localization}, in: \bibinfo{booktitle}{Proceedings of the 46th IEEE/ACM International Conference on Software Engineering}, pp. \bibinfo{pages}{1--12}.
%Type = Inproceedings
\bibitem[{Yang et~al.(2023)Yang, Hao, Quan, Yang, Zhao, Kuleshov and Wang}]{yang2023harnessing}
\bibinfo{author}{Yang, Q.}, \bibinfo{author}{Hao, Y.}, \bibinfo{author}{Quan, K.}, \bibinfo{author}{Yang, S.}, \bibinfo{author}{Zhao, Y.}, \bibinfo{author}{Kuleshov, V.}, \bibinfo{author}{Wang, F.}, \bibinfo{year}{2023}.
\newblock \bibinfo{title}{Harnessing biomedical literature to calibrate clinicians’ trust in ai decision support systems}, in: \bibinfo{booktitle}{Proceedings of the 2023 CHI Conference on Human Factors in Computing Systems}, pp. \bibinfo{pages}{1--14}.
%Type = Inproceedings
\bibitem[{Yang et~al.(2019)Yang, Steinfeld and Zimmerman}]{yang2019unremarkable}
\bibinfo{author}{Yang, Q.}, \bibinfo{author}{Steinfeld, A.}, \bibinfo{author}{Zimmerman, J.}, \bibinfo{year}{2019}.
\newblock \bibinfo{title}{Unremarkable ai: Fitting intelligent decision support into critical, clinical decision-making processes}, in: \bibinfo{booktitle}{Proceedings of the 2019 CHI Conference on Human Factors in Computing Systems}, pp. \bibinfo{pages}{1--11}.
%Type = Article
\bibitem[{Yin et~al.(2022)Yin, Tang, Cao, You, Wang and Alazab}]{yin2022knowledge}
\bibinfo{author}{Yin, J.}, \bibinfo{author}{Tang, M.}, \bibinfo{author}{Cao, J.}, \bibinfo{author}{You, M.}, \bibinfo{author}{Wang, H.}, \bibinfo{author}{Alazab, M.}, \bibinfo{year}{2022}.
\newblock \bibinfo{title}{Knowledge-driven cybersecurity intelligence: Software vulnerability coexploitation behavior discovery}.
\newblock \bibinfo{journal}{IEEE transactions on industrial informatics} \bibinfo{volume}{19}, \bibinfo{pages}{5593--5601}.
%Type = Inproceedings
\bibitem[{Zheng et~al.(2022)Zheng, Tang, Liu, Liu and Huang}]{zheng2022ux}
\bibinfo{author}{Zheng, Q.}, \bibinfo{author}{Tang, Y.}, \bibinfo{author}{Liu, Y.}, \bibinfo{author}{Liu, W.}, \bibinfo{author}{Huang, Y.}, \bibinfo{year}{2022}.
\newblock \bibinfo{title}{Ux research on conversational human-ai interaction: A literature review of the acm digital library}, in: \bibinfo{booktitle}{Proceedings of the 2022 CHI Conference on Human Factors in Computing Systems}, pp. \bibinfo{pages}{1--24}.
%Type = Inproceedings
\bibitem[{Zhou et~al.(2023)Zhou, Zhang, Luo, Parker and De~Choudhury}]{zhou2023synthetic}
\bibinfo{author}{Zhou, J.}, \bibinfo{author}{Zhang, Y.}, \bibinfo{author}{Luo, Q.}, \bibinfo{author}{Parker, A.G.}, \bibinfo{author}{De~Choudhury, M.}, \bibinfo{year}{2023}.
\newblock \bibinfo{title}{Synthetic lies: Understanding ai-generated misinformation and evaluating algorithmic and human solutions}, in: \bibinfo{booktitle}{Proceedings of the 2023 CHI Conference on Human Factors in Computing Systems}, pp. \bibinfo{pages}{1--20}.
%Type = Article
\bibitem[{Ziegele et~al.(2018)Ziegele, Weber, Quiring and Breiner}]{ziegele2018dynamics}
\bibinfo{author}{Ziegele, M.}, \bibinfo{author}{Weber, M.}, \bibinfo{author}{Quiring, O.}, \bibinfo{author}{Breiner, T.}, \bibinfo{year}{2018}.
\newblock \bibinfo{title}{The dynamics of online news discussions: Effects of news articles and reader comments on users’ involvement, willingness to participate, and the civility of their contributions}.
\newblock \bibinfo{journal}{Information, Communication \& Society} \bibinfo{volume}{21}, \bibinfo{pages}{1419--1435}.
%Type = Article
\bibitem[{Zolanvari et~al.(2021)Zolanvari, Yang, Khan, Jain and Meskin}]{zolanvari2021trust}
\bibinfo{author}{Zolanvari, M.}, \bibinfo{author}{Yang, Z.}, \bibinfo{author}{Khan, K.}, \bibinfo{author}{Jain, R.}, \bibinfo{author}{Meskin, N.}, \bibinfo{year}{2021}.
\newblock \bibinfo{title}{Trust xai: Model-agnostic explanations for ai with a case study on iiot security}.
\newblock \bibinfo{journal}{IEEE internet of things journal} .

\end{thebibliography}

\end{document}